\documentclass[11pt]{article}
\usepackage{appb,epsfig}

\usepackage{amsmath}
\usepackage{latexsym}

\def\lsim{\mathrel{\rlap{\raise 2.5pt \hbox{$<$}}\lower 2.5pt
\hbox{$\sim$}}}
\def\gsim{\mathrel{\rlap{\raise 2.5pt \hbox{$>$}}\lower 2.5pt
\hbox{$\sim$}}}
\def\thW{\theta_{\rm W}}
\def\GeV{{\rm GeV}}
\def\TeV{{\rm TeV}}
\def\dd{{\rm d}}
\newcommand{\veck}{{\pmb{k}}}
\def \sup{^{\vphantom{2}}}
\def\deg{\hbox{$^{\circ}$}}

\catcode`@=11
\def\citer{\@ifnextchar [{\@tempswatrue\@citexr}{\@tempswafalse\@citexr[]}}
 
\def\@citexr[#1]#2{\if@filesw\immediate\write\@auxout{\string\citation{#2}}\fi
  \def\@citea{}\@cite{\@for\@citeb:=#2\do
    {\@citea\def\@citea{--\penalty\@m}\@ifundefined
       {b@\@citeb}{{\bf ?}\@warning
       {Citation `\@citeb' on page \thepage \space undefined}}%
\hbox{\csname b@\@citeb\endcsname}}}{#1}}
\catcode`@=12

\begin{document}

\thispagestyle{empty}

\begin{flushright}
{\tt University of Bergen, Department of Physics}    \\[2mm]
{\tt Scientific/Technical Report No.1999-02}    \\[6mm]
{\tt ISSN 0803-2696} \\[9mm]
{hep-ph/9903301} \\[5mm]
{March 1999}       
\end{flushright}

\vspace*{44mm}

\begin{center}
{\bf \large
Higgs boson production in $e^+e^-$ and $e^-e^-$ collisions
}
\vspace{3mm}

P.~Osland

\vspace{3mm}

{\em Department of Physics, University of Bergen,\\
           All\'{e}gaten 55, N-5007 Bergen, Norway}

\end{center}
\newpage
\thispagestyle{empty}
\phantom{A}

\newpage

\pagenumbering{arabic}

\title{Higgs boson production in $e^+e^-$ and $e^-e^-$
collisions\thanks{Presented at Cracow 
Epiphany Conference on Electron-Positron Colliders,
5--10 January 1999, Cracow, Poland}
}
\author{Per Osland
\address{Department of Physics, University of Bergen,
N-5007 Bergen, Norway}
}
\maketitle

\begin{abstract}
When Higgs boson candidates will be found at future colliders,
it becomes imperative to determine their properties,
beyond the mass, production cross section and decay rates.
Other crucial properties are those related to the behaviour under 
$CP$ transformations, and the self-couplings.
This paper addresses the question of measurability of some of the
trilinear couplings of MSSM neutral Higgs bosons at a high-energy
$e^+e^-$ collider, and the possibilities of exploring
the Higgs boson $CP$ properties at $e^+e^-$ and $e^-e^-$ colliders.
\end{abstract}
\PACS{14.80.Cp, 12.60.Jv, 13.90.+i}

  
\section{Introduction}


The Higgs particle is expected to be discovered at the LHC,
if not already at LEP \cite{Zerwas}.
Current estimates from precision electroweak data \cite{LEP} 
suggest that it is rather light.
A light Higgs particle would be consistent with both the Standard
Model (SM) and the Minimal Supersymmetric Standard Model (MSSM).
A detailed measurement of its branching ratios should enable one 
to distinguish between these two most favoured models.

However, there is more to the MSSM Higgs sector than branching ratios.
For a complete analysis, one should also measure the trilinear and quartic
self-couplings, which in the MSSM are determined by the gauge couplings.

The measurability of couplings involving the light Higgs particle
was investigated by Djouadi, Haber and Zerwas \cite{DHZ}.
It was concluded that the trilinear couplings $\lambda_{Hhh}$ 
and $\lambda_{hhh}$, where $h$ and $H$ denote the two neutral, 
$CP$-even Higgs bosons, could be measured at a high-energy linear collider.
This early study neglected squark mixing, but, with some limitations,
the conclusion was confirmed also for the case of squark mixing \cite{OP98}.
A recent study also accounts for the dominant two-loop effects 
\cite{Muhlleitner}.

One should keep in mind that the Higgs sector might be more rich
than suggested by the MSSM \cite{GHKD}.
Thus, it would be most useful to establish the $CP$ properties of the Higgs 
particle from basic principles.
A straightforward method to determine the parity is to study the angular
distribution of the Higgs particle itself \cite{e-p,Accomando}.
A second approach makes use of the orientation of the plane spanned
by the fermions from the accompanying $Z$ boson in the Bjorken
process \cite{NelCha,Grzad,SkjOsl95}.

We also review the production of scalar Higgs-like particles 
in high-energy electron-electron collisions,
via the fusion of electroweak gauge bo\-sons.
The emphasis is on how to distinguish a
$CP$-even from a $CP$-odd Higgs particle \cite{Boe}.
Among the more significant differences, we find that
in the $CP$-odd case, the Higgs spectrum is much harder,
and the dependence of the total cross section on the product of the
polarizations of the two beams is much stronger, as compared with
the $CP$-even case.
We also briefly discuss parity violation, and the production of
charged Higgs bosons.

\section{Trilinear Higgs couplings}

Trilinear couplings of the neutral $CP$-even
Higgs bosons in the Minimal Supersymmetric 
Standard Model (MSSM) can be measured through
the multiple production of the lightest $CP$-even Higgs boson
at  high-energy $e^+  e^-$ colliders. 
The relevant production mechanisms are the production of 
the heavier $CP$-even Higgs boson via
$e^+e^- \rightarrow ZH$, in association with 
the $CP$-odd Higgs boson ($A$) in $e^+e^- \rightarrow AH$, 
or via the fusion process $e^+e^- \rightarrow \nu_e \bar\nu_e H$, 
with $H$ subsequently decaying through $H \rightarrow hh$.

The trilinear Higgs couplings that are of interest are 
$\lambda_{Hhh}$, $\lambda_{hhh}$, and $\lambda_{hAA}$, 
involving both the $CP$-even and $CP$-odd Higgs bosons.
The couplings $\lambda_{Hhh}$ and $\lambda_{hhh}$
are rather small with respect to the corresponding trilinear coupling
$\lambda_{hhh}^{\rm SM}$ in the SM (for a given mass 
of the lightest Higgs boson $m_h$), 
unless $m_h$ is close to the upper value (decoupling limit).
The coupling $\lambda_{hAA}$ remains small for all parameters.

We have considered the question of possible measurements
of the trilinear Higgs couplings $\lambda_{Hhh}$ and $\lambda_{hhh}$
\cite{OP98} of the MSSM \cite{HPN}
at a high-energy $e^+ e^-$ 
linear collider that will operate at an energy of 500~GeV 
with an integrated luminosity per year of  
${\mathcal L}_{\rm int} = 500~\mbox{fb}^{-1}$ \cite{NLC}.
In a later phase one may envisage an upgrade to an energy of 1.5~TeV.

The multiple production
of the light Higgs boson through Higgs\-strahlung of $H$, and 
through production of $H$ in association with the $CP$-odd Higgs 
boson can be used to 
extract the trilinear Higgs coupling $\lambda_{Hhh}$.
The non-resonant fusion mechanism for multiple
$h$ production, $e^+e^-\to \nu_e\bar\nu_e hh$, involves
two trilinear Higgs couplings, $\lambda_{Hhh}$ and $\lambda_{hhh}$,
and is useful for extracting $\lambda_{hhh}$.

In units of $gm_Z/(2\cos\thW)=(\sqrt{2}G_F)^{1/2}m_Z^2$,
the {\it tree-level} trilinear Higgs couplings involving $h$ are given by
\begin{eqnarray} 
\lambda_{Hhh}^0 & = & 2\sin2\alpha \sin(\beta + \alpha) - \cos 2\alpha
\cos(\beta + \alpha), \\
\label{Eq:lambda-Hhh0}
\lambda_{hhh}^0 & = & 3 \cos2\alpha \sin(\beta + \alpha), \\
\label{Eq:lambda-hhh0}
\lambda_{hAA}^0 & = & \cos2\beta \sin(\beta + \alpha),      
\label{Eq:lambda-hAA0}
\end{eqnarray}
with $\alpha$ the mixing angle in the $CP$-even Higgs sector, which 
is determined by the parameters of the $CP$-even
Higgs mass matrix. 

\begin{figure}[htb]
\refstepcounter{figure}
\label{Fig:lam-mh}
\addtocounter{figure}{-1}
\begin{center}
\setlength{\unitlength}{1cm}
\begin{picture}(12,7.5)
\put(-1.0,1)
{\mbox{\epsfysize=7.0cm\epsffile{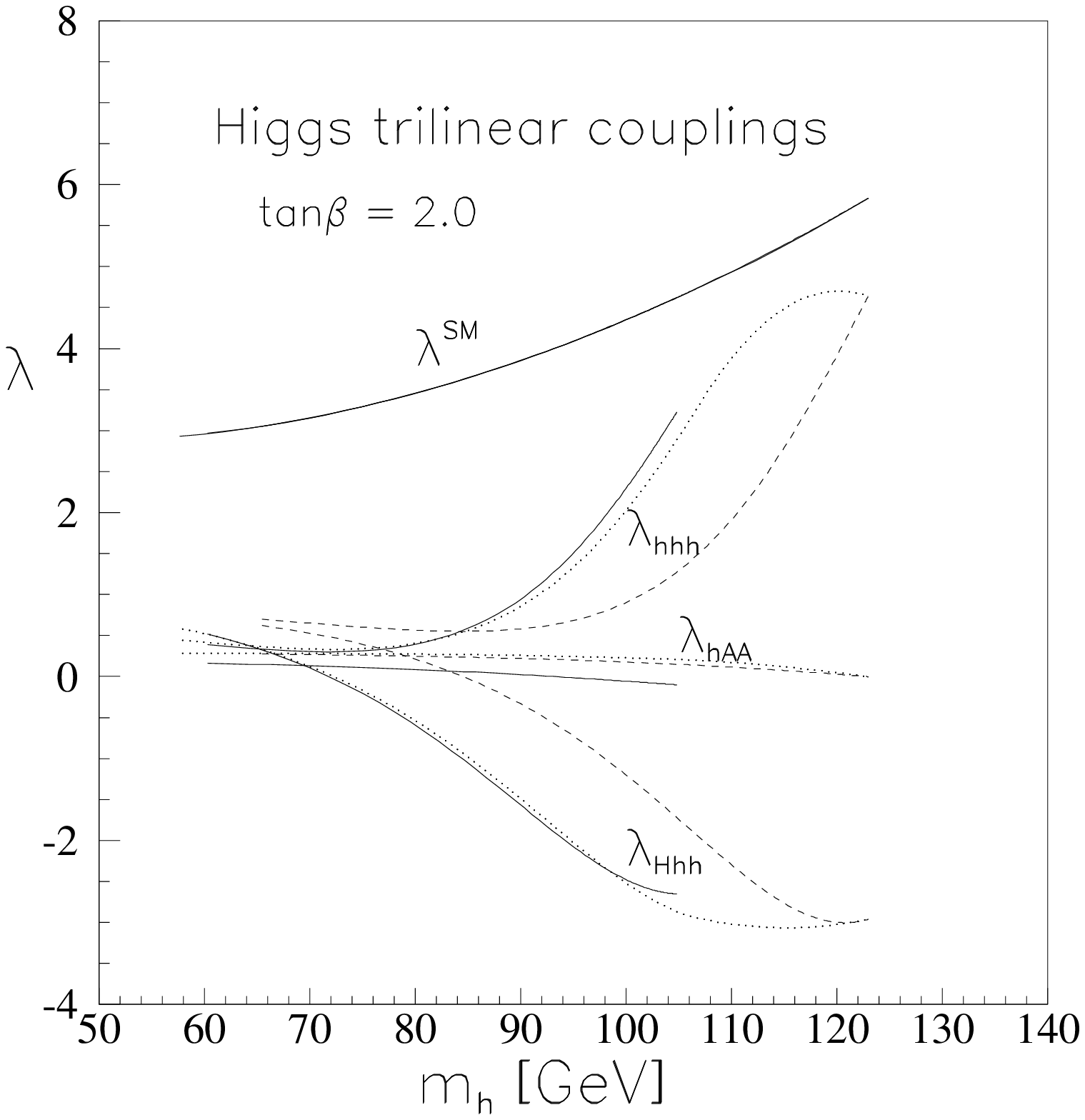}}
 \hspace*{-5mm}
 \mbox{\epsfysize=7.0cm\epsffile{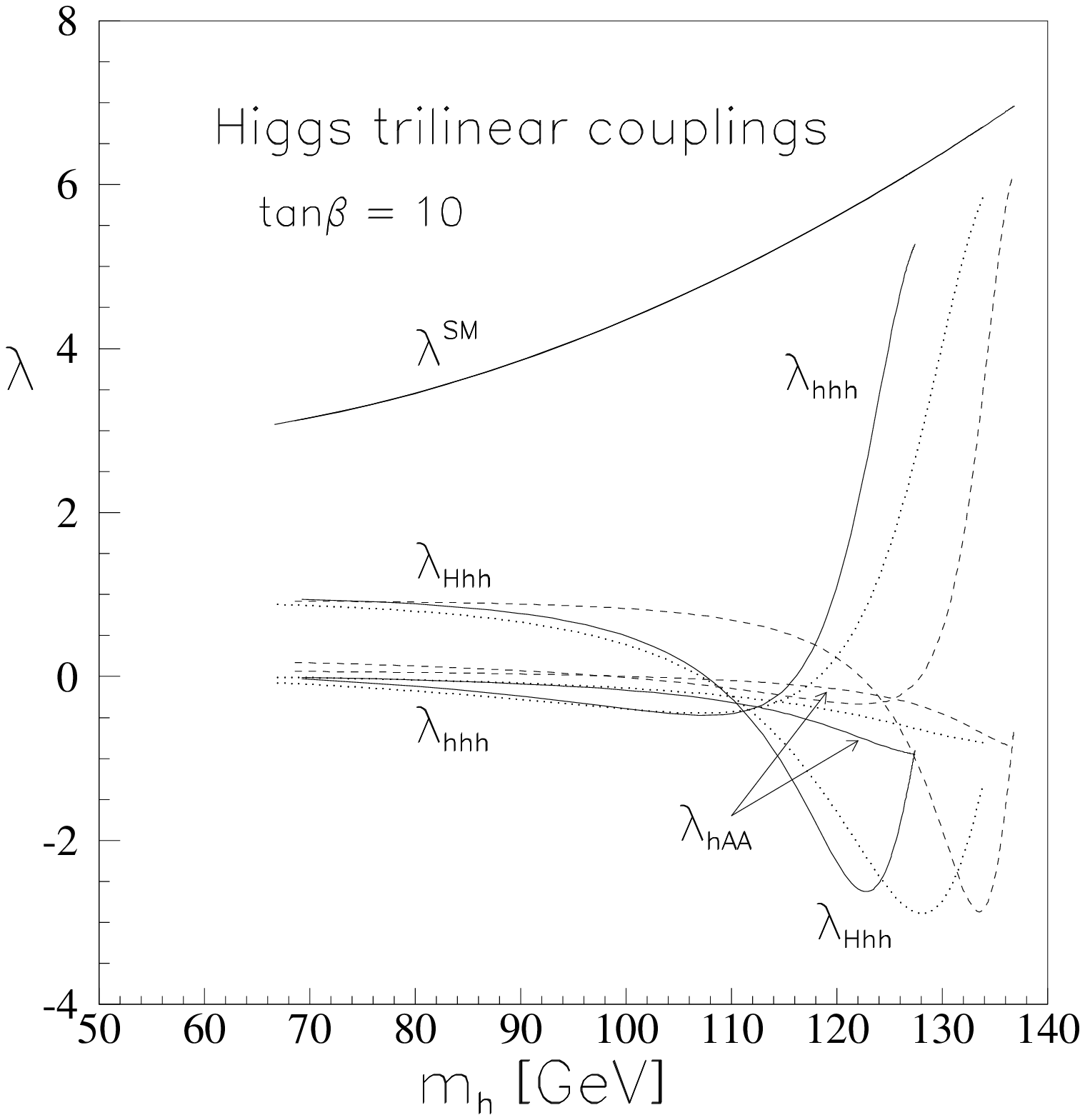}}}
\end{picture}
\vspace*{-8mm}
\caption{Trilinear Higgs couplings $\lambda_{Hhh}$, $\lambda_{hhh}$ and
$\lambda_{hAA}$ as functions of $m_h$ for 
$\tan\beta=2.0$ and $\tan\beta=10.0$.
Each coupling is shown for $\tilde m=1~\TeV$, and for
three cases of the mixing parameters:
no mixing ($A=0$, $\mu=0$, solid),
mixing with $A=1$~TeV and $\mu=-1$~TeV (dotted),
as well as 
$A=1$~TeV and $\mu=1$~TeV (dashed).}
\end{center}
\end{figure}

We include one-loop radiative corrections
\cite{ERZ1,BBSP} to the Higgs sector in the effective potential
approximation. In particular, we take into account \cite{OP98}
the parameters $A$ and $\mu$, the soft supersymmetry
breaking trilinear parameter and the bilinear Higgs(ino)  
parameter in the superpotential.
These parameters enter through the stop masses,
\begin{equation}
m_{\tilde t_{1,2}}^2   =  m_t^2 + \tilde m^2 \pm 
m_t(A + \mu\cot\beta)
\end{equation}
which again enter through the radiative corrections to the Higgs masses
as well as to the Higgs trilinear couplings.
The dominant one-loop radiative corrections are proportional
to $(m_t/m_W)^4$, multiplying functions depending on the squark masses
\cite{ERZ1,BBSP}.

The trilinear couplings depend significantly on $m_A$,
and thus also on $m_h$. This is shown in Fig.~\ref{Fig:lam-mh},
where we compare $\lambda_{Hhh}$, $\lambda_{hhh}$ and $\lambda_{hAA}$
for three different values of $\tan\beta$,
and the SM quartic coupling $\lambda^{\rm SM}$ (which also includes
one-loop radiative corrections \cite{SirZuc}).
For a given value of $m_h$, the values of these
couplings significantly depend on the soft supersymmetry-breaking 
trilinear parameter $A$, as well as on $\mu$.

As is clear from Fig.~\ref{Fig:lam-mh},
at low values of $m_h$, the MSSM trilinear couplings are rather small.
For some value of $m_h$ the couplings $\lambda_{Hhh}$ and $\lambda_{hhh}$
start to increase in magnitude, whereas $\lambda_{hAA}$ remains small.
The values of $m_h$ at which they start becoming significant
depend crucially on $\tan\beta$.

To sum up the behaviour of the trilinear couplings, we note that
$\lambda_{Hhh}$ and  $\lambda_{hhh}$ are small for 
$m_h \lsim 100$--120~GeV, depending on the value of $\tan\beta$. 
However, as $m_h$ approaches its maximum value, 
which requires $m_A \gsim 200$~GeV, these trilinear couplings become 
reasonably large.

\section{Production mechanisms}

Different mechanisms for multiple production of the MSSM 
Higgs bosons in $e^+ e^-$ collisions have been discussed by DHZ.
The dominant mechanism for the production of multiple  
$CP$-even light Higgs bosons is through the mechanisms 
\begin{eqnarray}
\left. \begin{array}{ccc}
e^+e^- & \rightarrow & ZH,AH \\ 
e^+e^- & \rightarrow & \nu_e \bar \nu_e H
\end{array}
\right\}, \qquad H \rightarrow hh, \label{Eq:res-Hhh} 
\end{eqnarray}
shown in Fig.~\ref{Fig:Feynman-resonant}. 
The heavy Higgs boson $H$ can be produced
by $H$-strahlung, in association with $A$, 
and by the resonant $WW$ fusion mechanism. 
All the diagrams of Fig.~\ref{Fig:Feynman-resonant} involve the
trilinear coupling $\lambda_{Hhh}$.

\begin{figure}[htb]
\refstepcounter{figure}
\label{Fig:Feynman-resonant}
\addtocounter{figure}{-1}
\begin{center}
\setlength{\unitlength}{1cm}
\begin{picture}(12,7.5)
\put(-0.5,-11.5)
{\mbox{\epsfxsize=16cm\epsffile{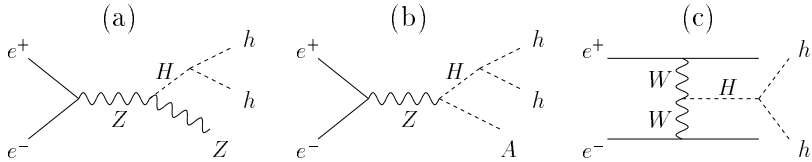}}}
\end{picture}
\vspace*{-50mm}
\caption{Feynman diagrams for the resonant production
of $hh$ final states in $e^+ e^-$ collisions.}
\end{center}
\end{figure}
\vspace*{-5mm}

A background to (\ref{Eq:res-Hhh}) comes from the production of the 
pseudoscalar $A$ in association with $h$ and its subsequent decay to $hZ$
\begin{equation}
e^+e^- \rightarrow hA, \qquad A \rightarrow hZ, \label{Eq:bck-hA}
\end{equation}
leading to $Zhh$ final states.
A further mechanism for $hh$ production is double Higgs-strahlung 
in the continuum with a $Z$ boson in the final state,
\begin{equation}
e^+e^-  \rightarrow Z^* \rightarrow Zhh. \label{Eq:Zstar}
\end{equation} 

There is also a mechanism of multiple production of the lightest Higgs
boson through non-resonant $WW$ fusion in the continuum:
\begin{equation}
e^+e^-  \rightarrow \bar \nu_e \nu_e W^* W^* \rightarrow 
\bar \nu_e \nu_e hh, \label{Eq:WW-fusion}
\end{equation}
as shown in Fig.~\ref{Fig:Feynman-nonres-WW}.
\begin{figure}[htb]
\refstepcounter{figure}
\label{Fig:Feynman-nonres-WW}
\addtocounter{figure}{-1}
\begin{center}
\setlength{\unitlength}{1cm}
\begin{picture}(12,7)
\put(0,-9.5)
{\mbox{\epsfxsize=14cm\epsffile{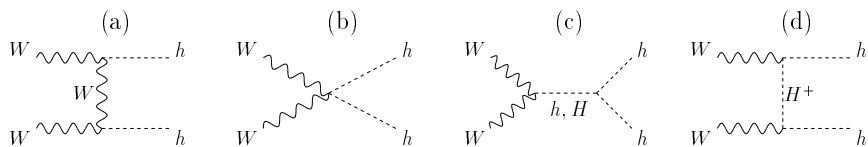}}}
\end{picture}
\vspace*{-45mm}
\caption{Feynman diagrams for the non-resonant $WW$ fusion
mechanism for the production of $hh$ states in $e^+ e^-$ collisions.}
\end{center}
\end{figure}
\vspace*{-5mm}

It is important to note that all the diagrams of  
Fig.~\ref{Fig:Feynman-resonant}
involve the trilinear coupling $\lambda_{Hhh}$ only. 
In contrast,
the non-resonant analogues of Figs.~\ref{Fig:Feynman-resonant}a,
\ref{Fig:Feynman-resonant}b and \ref{Fig:Feynman-resonant}c
(or \ref{Fig:Feynman-nonres-WW}c) involve both
the trilinear Higgs couplings $\lambda_{Hhh}$ and $\lambda_{hhh}$.

\subsection{Higgs-strahlung and associated production of $H$}
The dominant source for the production of
multiple light Higgs bosons in $e^+ e^-$ collisions is through 
the production of the heavier $CP$-even Higgs boson $H$ either via
Higgs-strahlung or in association with $A$,  
followed, if kinematically allowed, by the decay 
$H \rightarrow hh$.

\begin{figure}[htb]
\refstepcounter{figure}
\label{Fig:sigma-500-1500}
\addtocounter{figure}{-1}
\begin{center}
\setlength{\unitlength}{1cm}
\begin{picture}(12,7.8)
\put(-1.0,1.5)
{\mbox{\epsfysize=7.0cm\epsffile{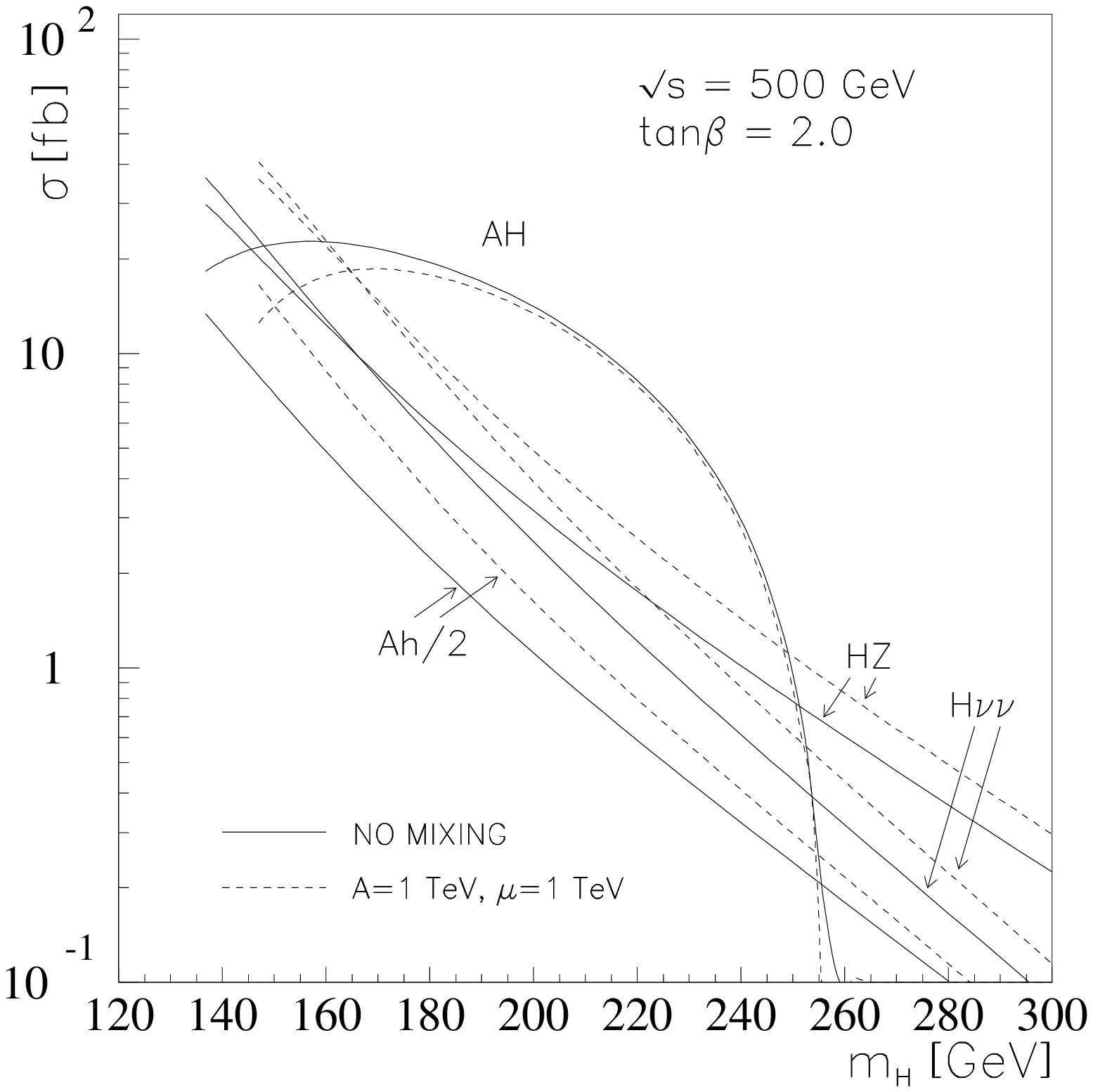}}
 \mbox{\epsfysize=7.0cm\epsffile{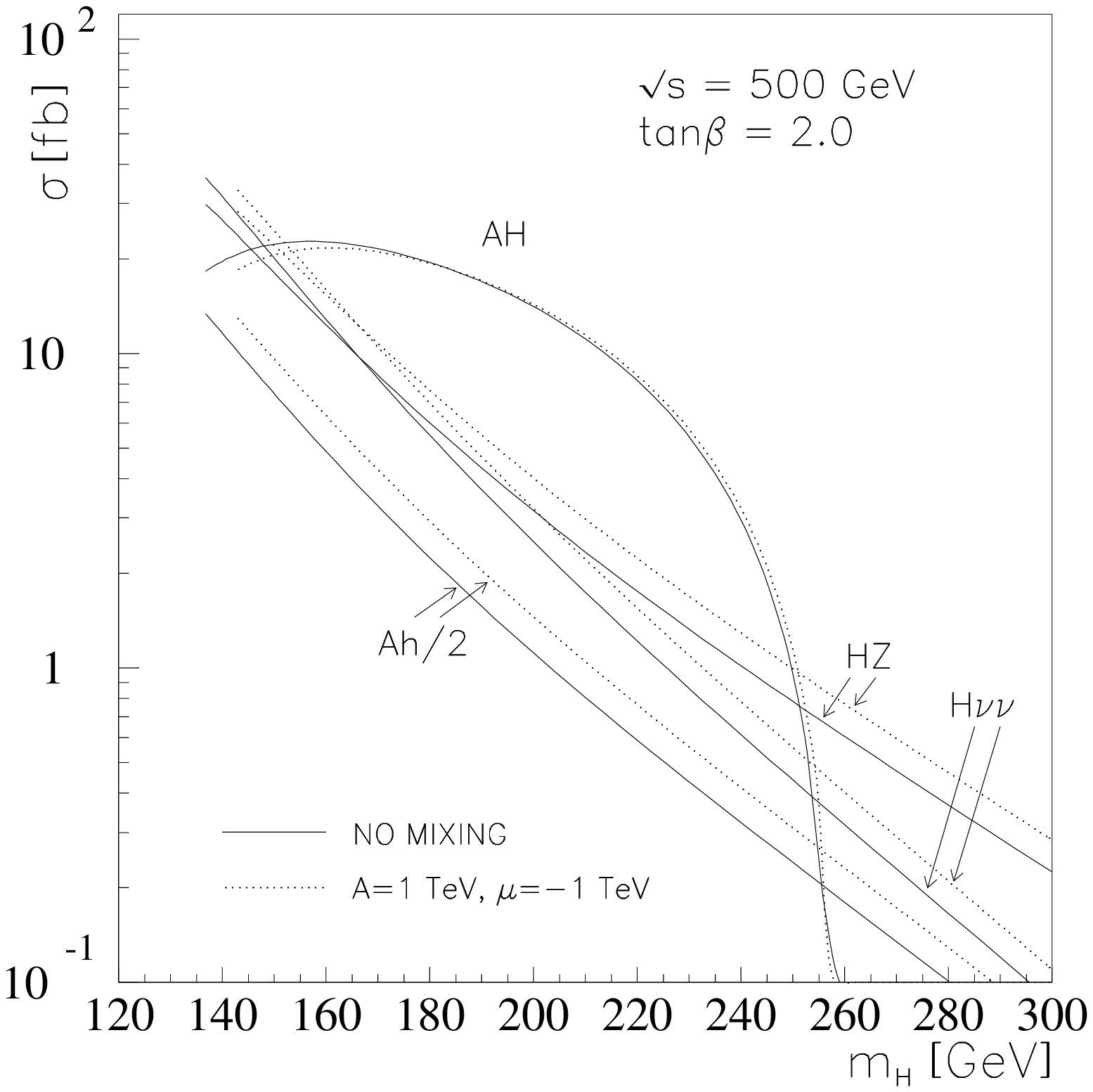}}}
\end{picture}
\vspace*{-15mm}
\caption{Cross sections for the production of the heavy Higgs
boson $H$ in $e^+ e^-$ collisions,
and for the background process in which $Ah$ is produced. 
Solid curves are for no mixing, $A=0$, $\mu=0$. 
Dashed and dotted curves refer to mixing.}
\end{center}
\end{figure}

In Fig.~\ref{Fig:sigma-500-1500} we plot the relevant cross sections 
\cite{PocZsi,GETAL} for the 
$e^+e^-$ centre-of-mass energy $\sqrt s = 500~\GeV$,
as functions of the Higgs mass $m_H$ and for $\tan\beta = 2.0$. 
For a fixed value of $m_H$, there is seen to be a significant
sensitivity to the squark mixing parameters $\mu$ and $A$.
We have here taken $\tilde m=1~\TeV$, a value which is adopted
throughout, except where otherwise specified.

A measurement of the decay rate $H \rightarrow hh$ directly
yields $\lambda_{Hhh}^2$.
But this is possible only if the decay is kinematically
allowed, and the branching ratio is sizeable (but not too close to unity).
In Fig.~\ref{Fig:BR-H-A} we show the branching ratios (at $\tan\beta=2$)
for the main decay modes of the heavy $CP$-even Higgs boson 
as a function of the $H$ mass \cite{DKZ1}.
Apart from the $hh$ decay mode, the other important decay modes 
are $H \rightarrow WW^*$, $ZZ^*$.
For increasing values of $\tan\beta$ (but fixed $m_h$), 
the $Hhh$ coupling gradually gets weaker (Fig.~\ref{Fig:lam-mh}),
and hence the prospects for measuring $\lambda_{Hhh}$ diminish.
Also, the decay rates can change significantly with $\tilde m$,
the over-all squark mass scale (see Fig.~\ref{Fig:BR-H-A}).

\begin{figure}[hb]
\refstepcounter{figure}
\label{Fig:BR-H-A}
\addtocounter{figure}{-1}
\begin{center}
\setlength{\unitlength}{1cm}
\begin{picture}(12,7.8)
\put(-1.0,1.7)
{\mbox{\epsfysize=7.0cm\epsffile{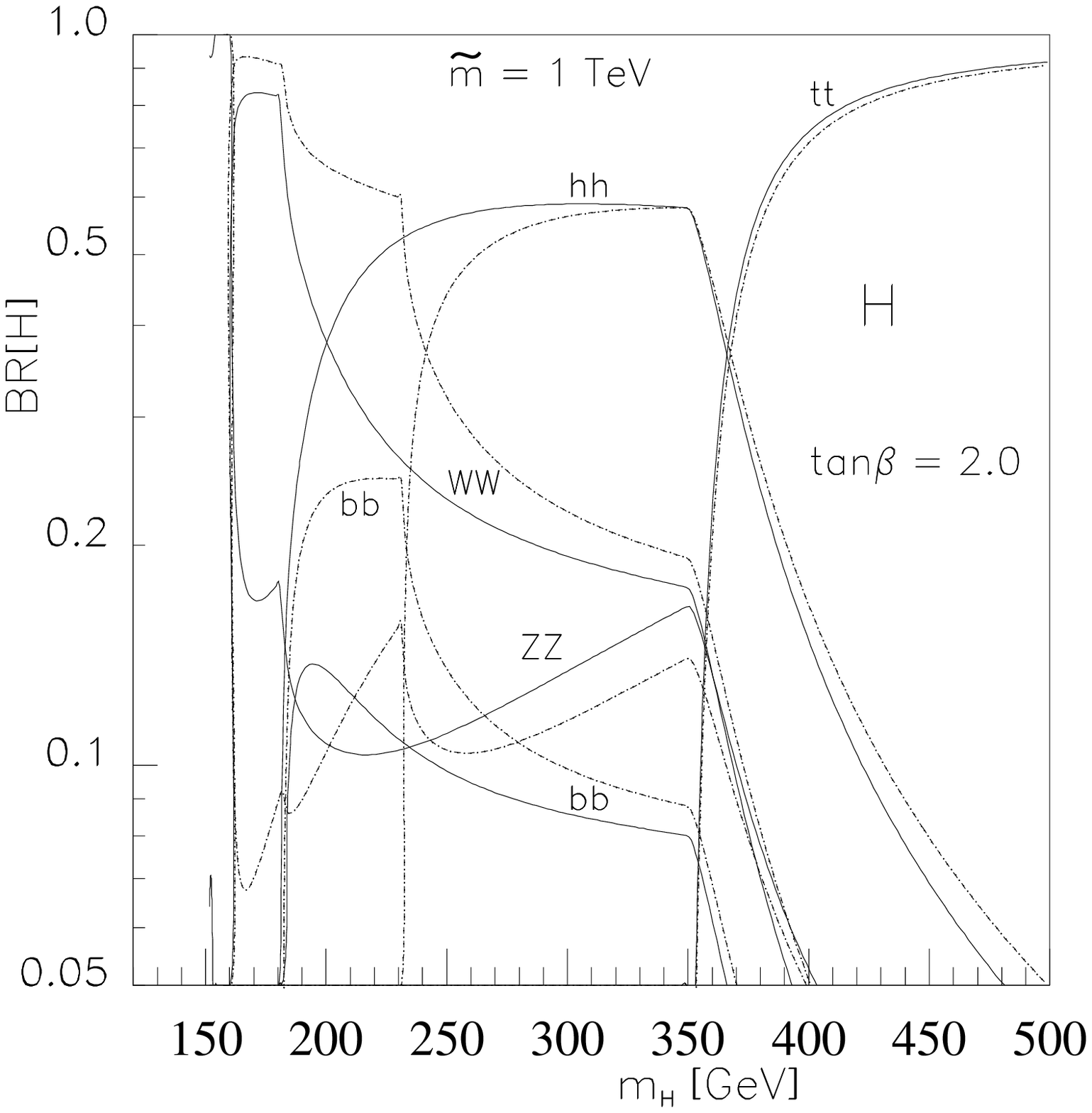}}
 \mbox{\epsfysize=7.0cm\epsffile{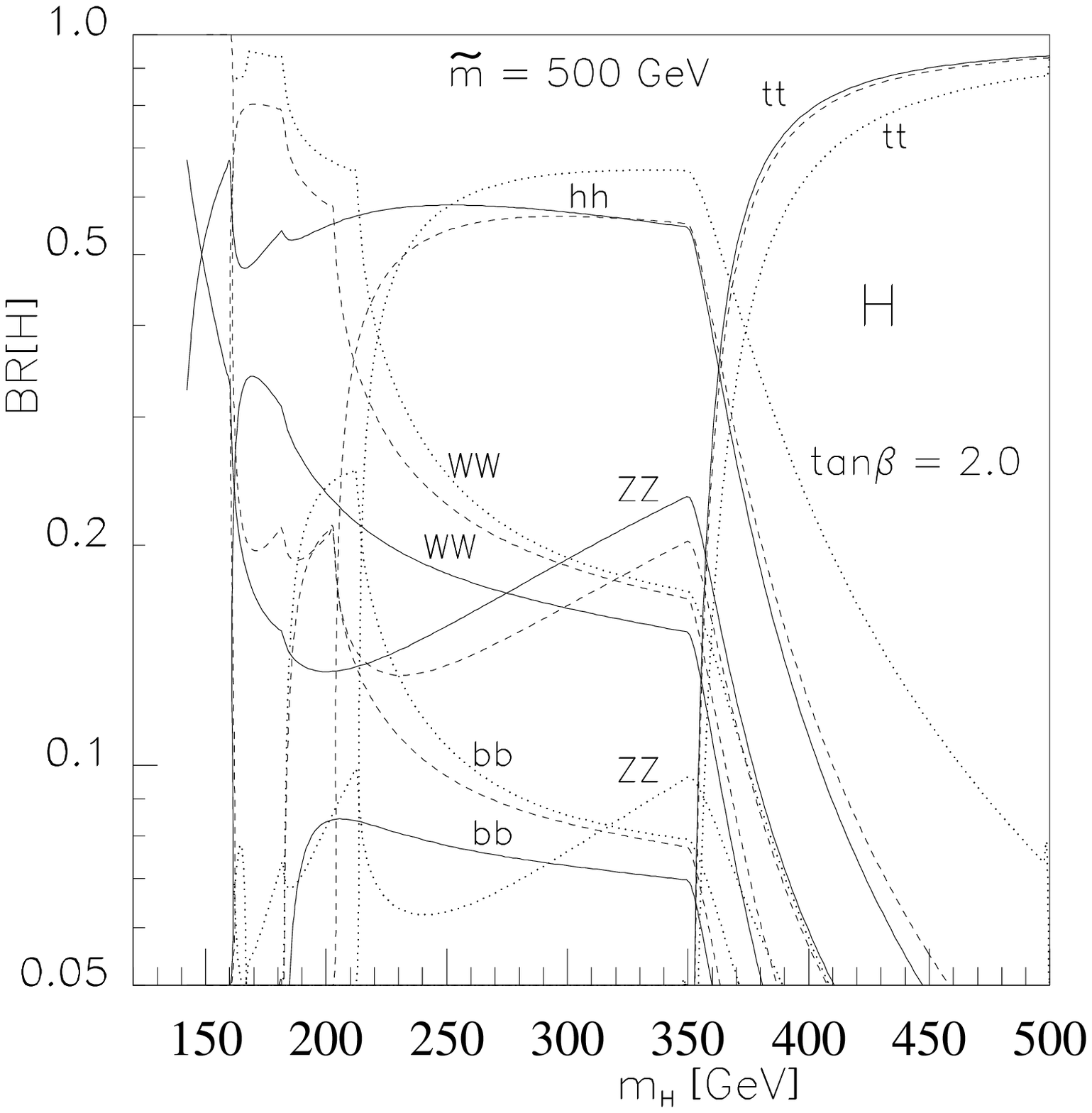}}}
\end{picture}
\vspace*{-15mm}
\caption{Branching ratios for the decay modes of the $CP$-even
heavy Higgs boson $H$, for $\tan\beta = 2.0$ and 
$\tilde m$ equal to 1~TeV or 500~GeV, 
as indicated.
Solid curves are for no mixing, $A=0$, $\mu=0$.
For $\tilde m=1$~TeV, the dashed curves refer to $A=1$~TeV and $\mu=1$~TeV,
whereas for $\tilde m=500$~GeV, the dashed (dotted) curves refer to 
$A=500$~GeV (800~GeV) and $\mu=1$~TeV (800~GeV).}
\end{center}
\end{figure}

\begin{figure}[htb]
\refstepcounter{figure}
\label{Fig:hole-1000}
\addtocounter{figure}{-1}
\begin{center}
\setlength{\unitlength}{1cm}
\begin{picture}(12,7.5)
\put(-1.0,1.5)
{\mbox{\epsfysize=7.0cm\epsffile{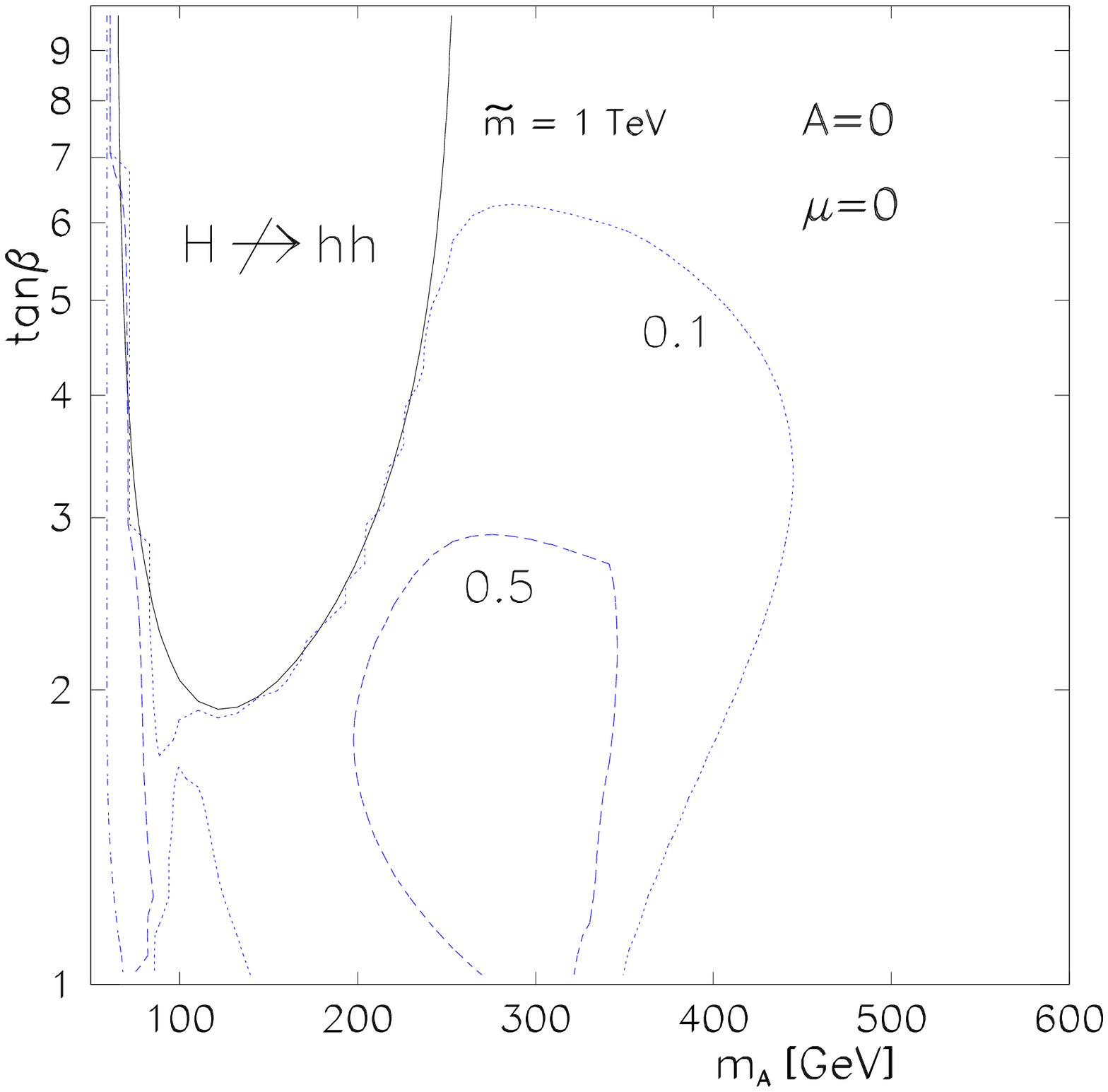}}
 \mbox{\epsfysize=7.0cm\epsffile{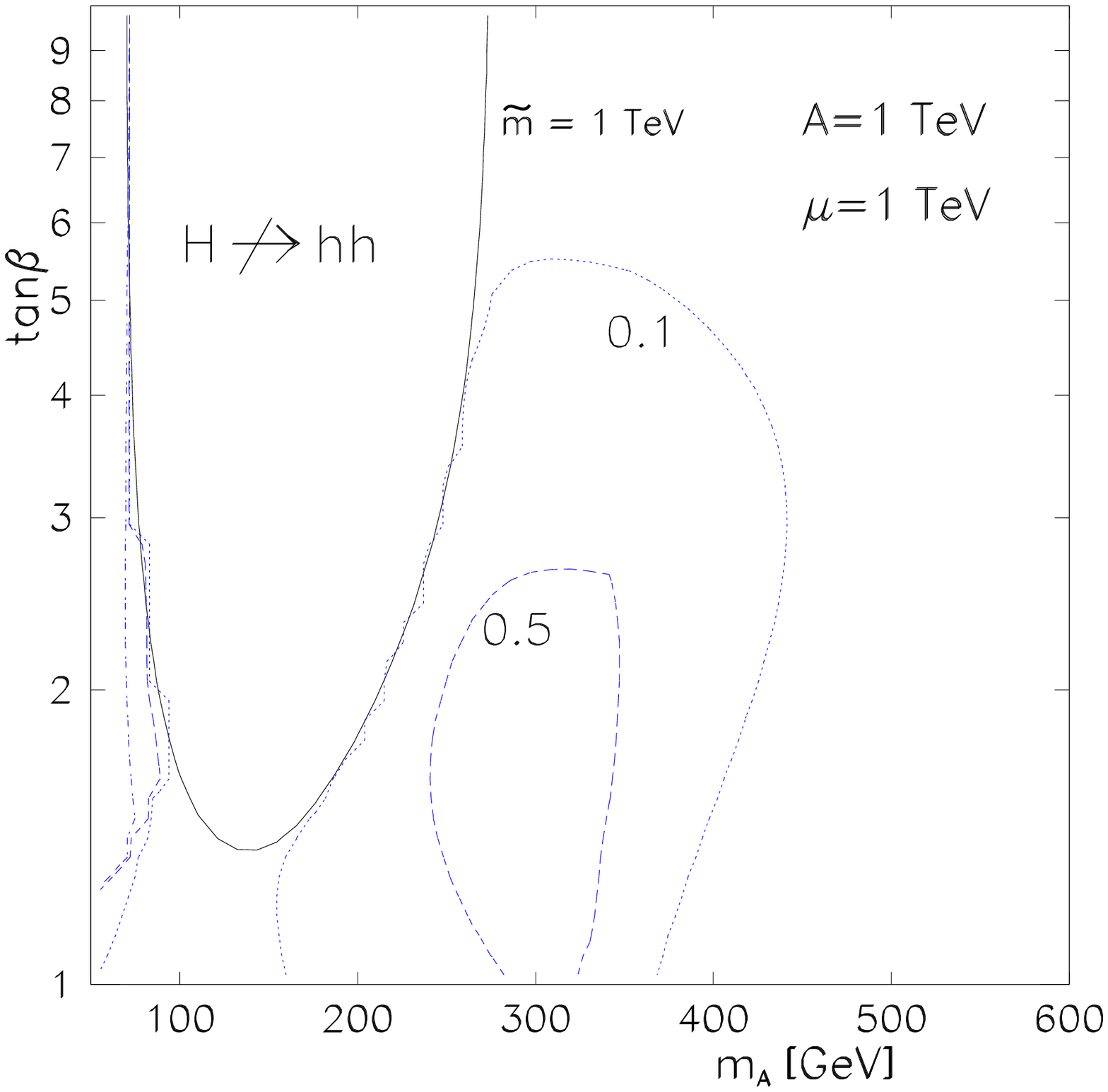}}}
\end{picture}
\vspace*{-12mm}
\caption{The region in the $m_A$--$\tan\beta$ plane where the decay
$H\to hh$ is kinematically {\em forbidden} is indicated by a solid 
line contour.
Also given are contours at which the branching ratio equals 0.1 (dotted),
0.5 (dashed) and 0.9 (dash-dotted, far left).}
\end{center}
\end{figure}

\begin{figure}[htb]
\refstepcounter{figure}
\label{Fig:hole-500}
\addtocounter{figure}{-1}
\begin{center}
\setlength{\unitlength}{1cm}
\begin{picture}(12,7.5)
\put(-1.0,1.5)
{\mbox{\epsfysize=7.0cm\epsffile{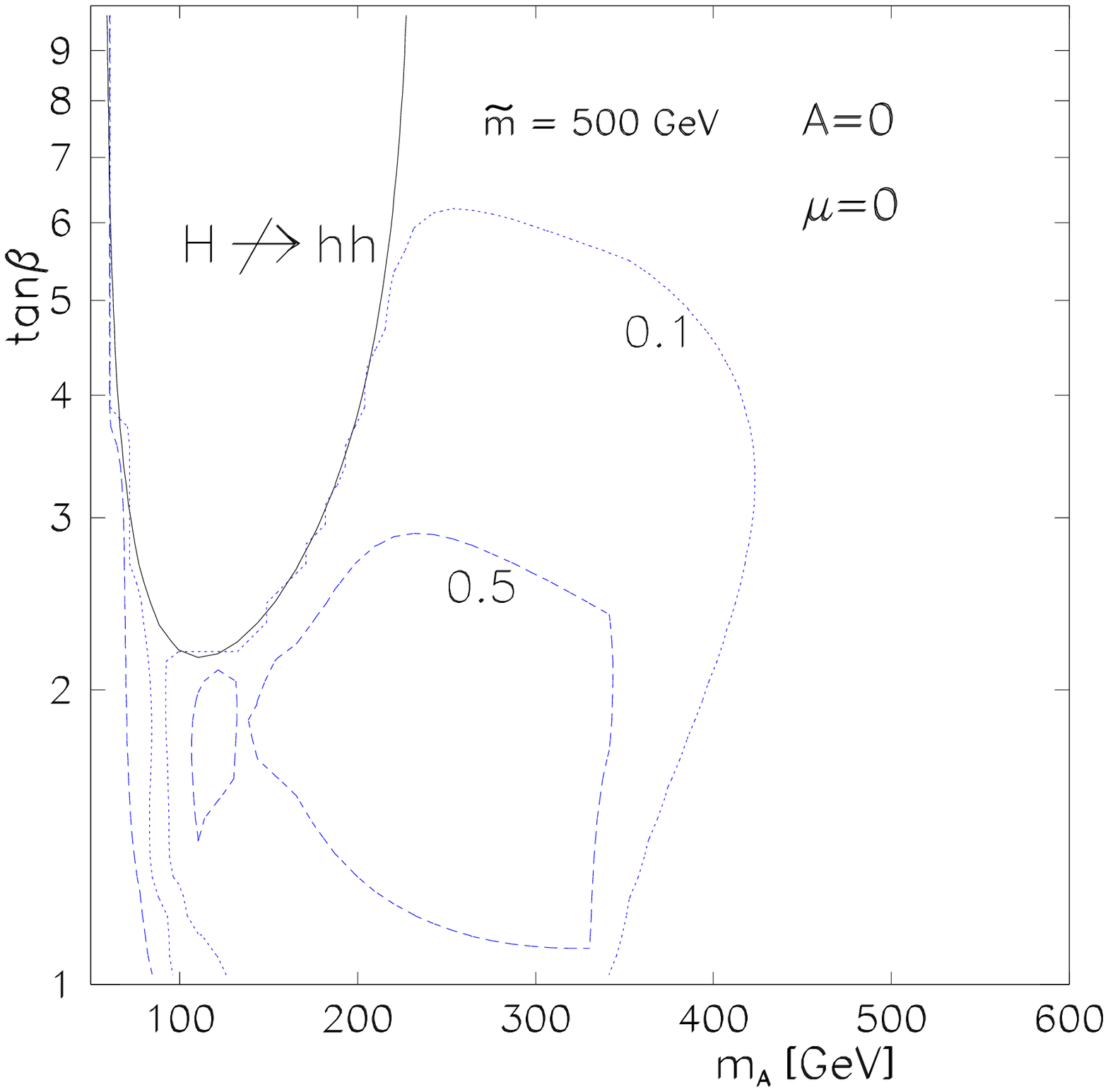}}
 \mbox{\epsfysize=7.0cm\epsffile{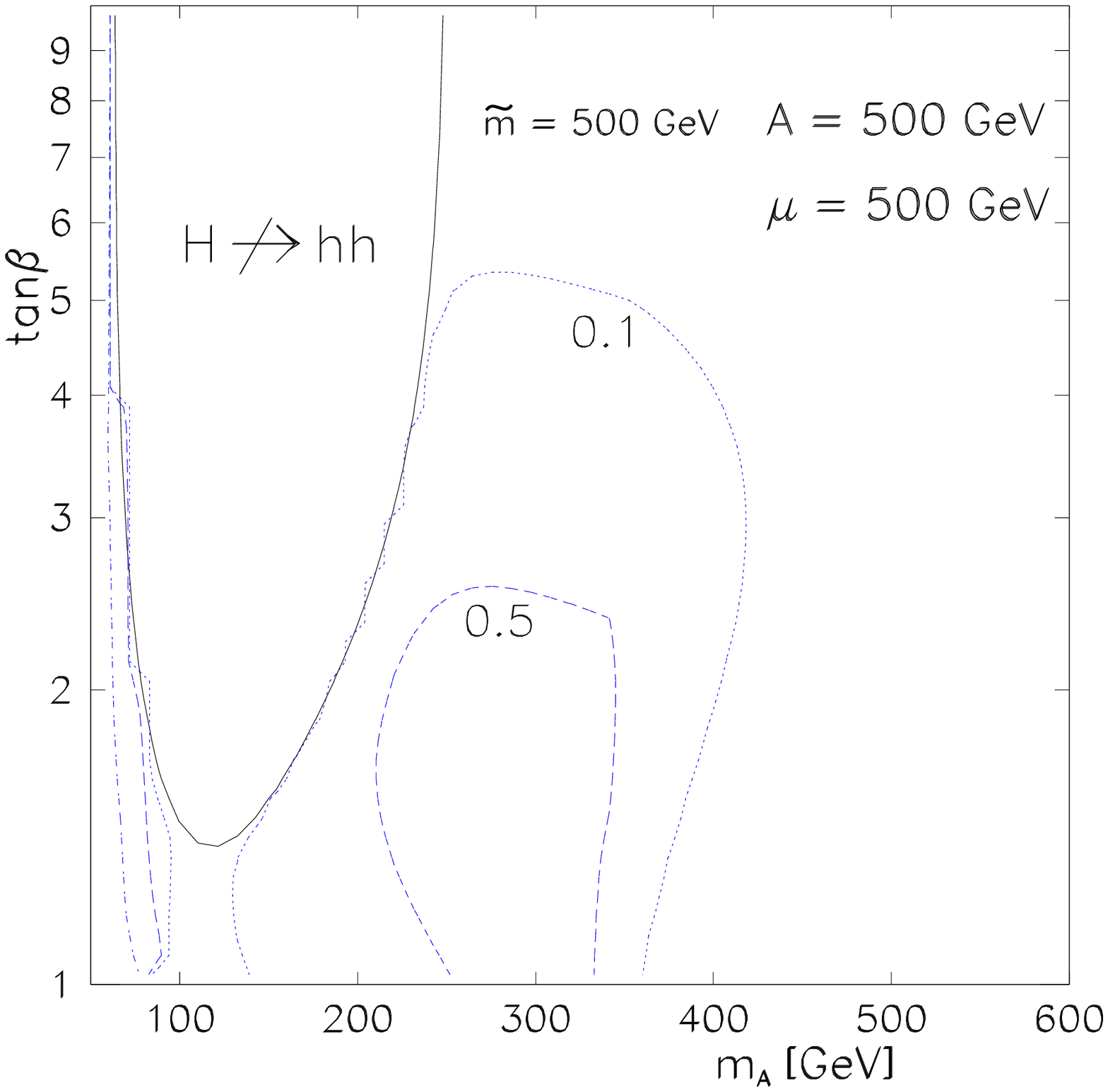}}}
\end{picture}
\vspace*{-12mm}
\caption{Similar to Fig.~\ref{Fig:hole-1000}, for squark mass parameter
$\tilde m=500~\GeV$.}
\end{center}
\end{figure}

There is a sizeable region in the $m_A$--$\tan\beta$ plane
where the decay $H\to hh$ is kinematically forbidden,
shown in Figs.~\ref{Fig:hole-1000} and \ref{Fig:hole-500},
as an egg-shaped region at the upper left.
The boundary of the region depends 
crucially on the precise Higgs mass values.
This is illustrated by comparing two cases of mixing parameters
$A$ and $\mu$ at each of two values of the squark mass parameter
$\tilde m$.
We also display the regions where the $H\to hh$
branching ratio is in the range 0.1--0.9.
Obviously, in the forbidden region, the $\lambda_{Hhh}$ cannot be
determined from resonant production.

\subsection{Double Higgs-strahlung}
As discussed above,
for small and moderate values of $\tan\beta$, a study of decays
of the heavy $CP$-even Higgs boson $H$ provides a means of determining
the triple-Higgs coupling $\lambda_{Hhh}$.
For the purpose of extracting the coupling $\lambda_{hhh}$, 
non-resonant processes
involving two-Higgs ($h$) final states must be considered.
The $Zhh$ final states produced in the non-resonant double 
Higgs-strahlung $e^+e^- \rightarrow Zhh$, and whose cross section
involves the coupling $\lambda_{hhh}$, 
could provide one possible opportunity.
\begin{figure}[htb]
\refstepcounter{figure}
\label{Fig:sig-Zll-2}
\addtocounter{figure}{-1}
\begin{center}
\setlength{\unitlength}{1cm}
\begin{picture}(12,7.8)
\put(-1.0,1.5)
{\mbox{\epsfysize=7.0cm\epsffile{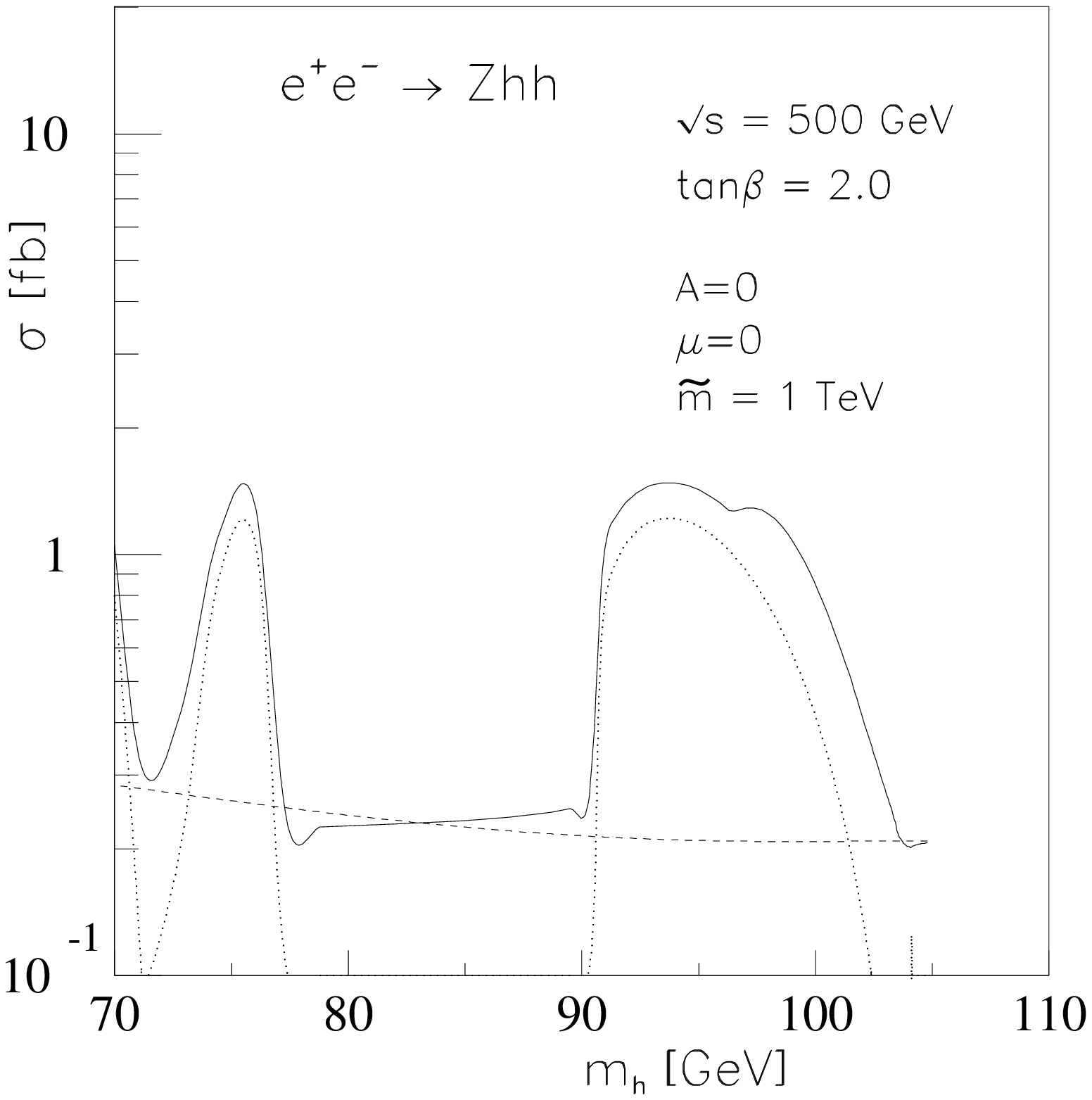}}
 \mbox{\epsfysize=7.0cm\epsffile{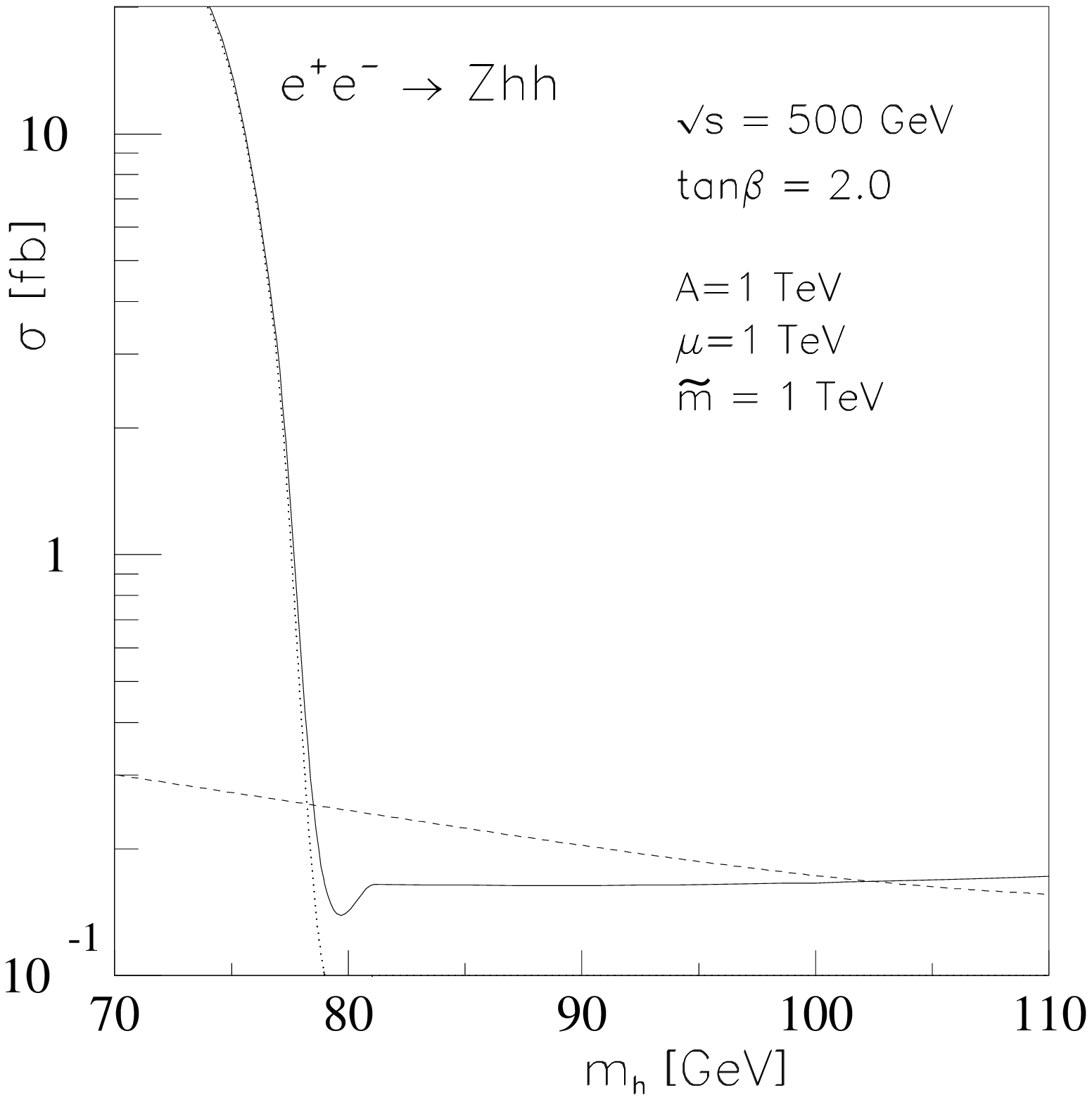}}}
\end{picture}
\vspace*{-12mm}
\caption{Cross section $\sigma(e^+e^-\to Zhh)$ as a function of $m_h$.
The dotted curve is the resonant production,
the dashed curve gives the decoupling limit \cite{HABER1}.}
\end{center}
\end{figure}

However, the non-resonant contribution to the $Zhh$ cross section
is rather small, as is shown 
in Fig.~\ref{Fig:sig-Zll-2} for $\sqrt s=500~\GeV$, $\tan\beta=2$,
and $\tilde m = 1~\TeV$.
In this case, the cross section is rather different for the two
sets of mixing parameters shown \cite{OP98}.
In the case of no mixing, there is a broad minimum from $m_h\simeq78$
to 90~GeV, followed by an enhancement around $m_h\sim90$--100~GeV.
This structure is in part due to the fact that the decay
$H\to hh$ is kinematically forbidden in the region 
$m_h\simeq78$--90~GeV, see Figs.~\ref{Fig:hole-1000} 
and \ref{Fig:hole-500} (this coincides with 
the opening up of the channel $H\to WW$), 
followed by an increase of the trilinear couplings.

Since the non-resonant part of the cross section, which depends on 
$\lambda_{hhh}$, is rather small, this channel is not suitable for
a determination of $\lambda_{hhh}$  \cite{DHZ}.
In the case of large squark mixing, the cross section can be considerably
larger \cite{OP98}, but only at Higgs masses which are essentially
ruled out\footnote{The LEP experiments have obtained strong
lower bounds on the mass of the lightest Higgs boson, and are
beginning to rule out significant parts of the small-$\tan\beta$ 
parameter space.
ALEPH finds a lower limit of $m_h>72.2$~GeV, irrespective of $\tan\beta$,
and a limit of $\sim 88$~GeV for $1<\tan\beta\lsim2$ \cite{ALEPH98}.}.
At higher values of $\tan\beta$, the cross section is even smaller.
For lower values of the squark mass parameter $\tilde m$,
the cross section can be larger, but again at Higgs masses
which are ruled out.

\subsection{Fusion mechanism for multiple-$h$ production}
A two-Higgs ($hh$) final state
can also result from the $WW$ fusion mechanism in $e^+ e^-$ collisions.
There is a resonant contribution (through $H$) and a non-resonant one. 

The resonant $WW$ fusion cross section for 
$e^+e^- \rightarrow H\bar\nu_e\nu_e$ \cite{DHKMZ}
is plotted in Fig.~\ref{Fig:sigma-500-1500} 
for the centre-of-mass energy $\sqrt s = 500$ GeV, 
and for $\tan\beta = 2.0$, as a function of $m_H$.  

Besides the resonant $WW$ fusion mechanism for the multiple
production of $h$ bosons, there is also a non-resonant $WW$ 
fusion mechanism:
\begin{equation}
\label{Eq:WW-nonres}
e^+e^-\to\nu_e\bar\nu_e hh,
\end{equation}
through which the same final state of two $h$ bosons can be produced.
The cross section for this process (see Fig.~\ref{Fig:Feynman-nonres-WW}),
can be written in the effective $WW$ approximation as
a $WW$ cross section, at invariant energy squared $\hat s=xs$, 
folded with the $WW$ ``luminosity'' \cite{CDCG}. Thus,
\begin{equation}
\label{Eq:sigWW-nonres}
\sigma(e^+e^-\to\nu_e\bar\nu_e hh)
=\int_\tau^1\dd x\, \frac{\dd L}{\dd x}\, \hat\sigma\sup_{WW}(x),
\end{equation}
where $\tau=4m_h^2/s$, and
\begin{equation}
\frac{\dd L(x)}{\dd x}
=\frac{G_{\rm F}^2m_W^4}{2}\,\left(\frac{1}{2\pi^2}\right)^2
\frac{1}{x}\biggl\{(1+x)\log\frac{1}{x}-2(1-x)\biggr\}.
\end{equation}

The $WW$ cross section receives contributions from several amplitudes,
according to the diagrams (a)--(d) 
in Fig.~\ref{Fig:Feynman-nonres-WW}, only one of which
is proportional to $\lambda_{hhh}$.
We have evaluated these contributions \cite{OP98}, 
following the approach of Ref.~\cite{AMP}, ignoring transverse
momenta everywhere except in the $W$ propagators.
Our approach also differs from that of \cite{DHZ} in that 
we do not project out 
the longitudinal degrees of freedom of the intermediate $W$ bosons.

\begin{figure}[htb]
\refstepcounter{figure}
\label{Fig:sig-WW-2}
\addtocounter{figure}{-1}
\begin{center}
\setlength{\unitlength}{1cm}
\begin{picture}(12,7.8)
\put(-1,1.5)
{\mbox{\epsfysize=7.0cm\epsffile{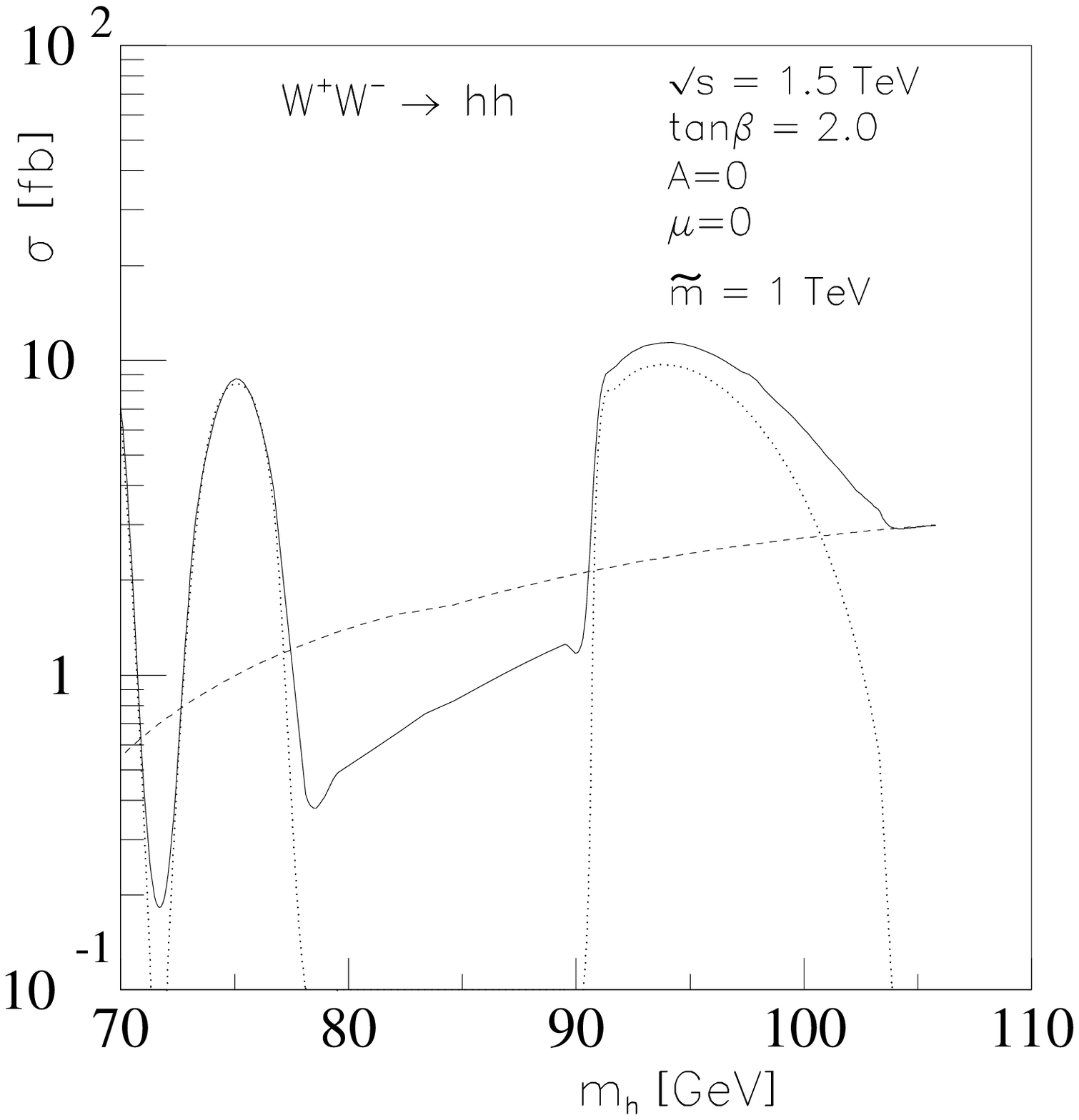}}
 \mbox{\epsfysize=7.0cm\epsffile{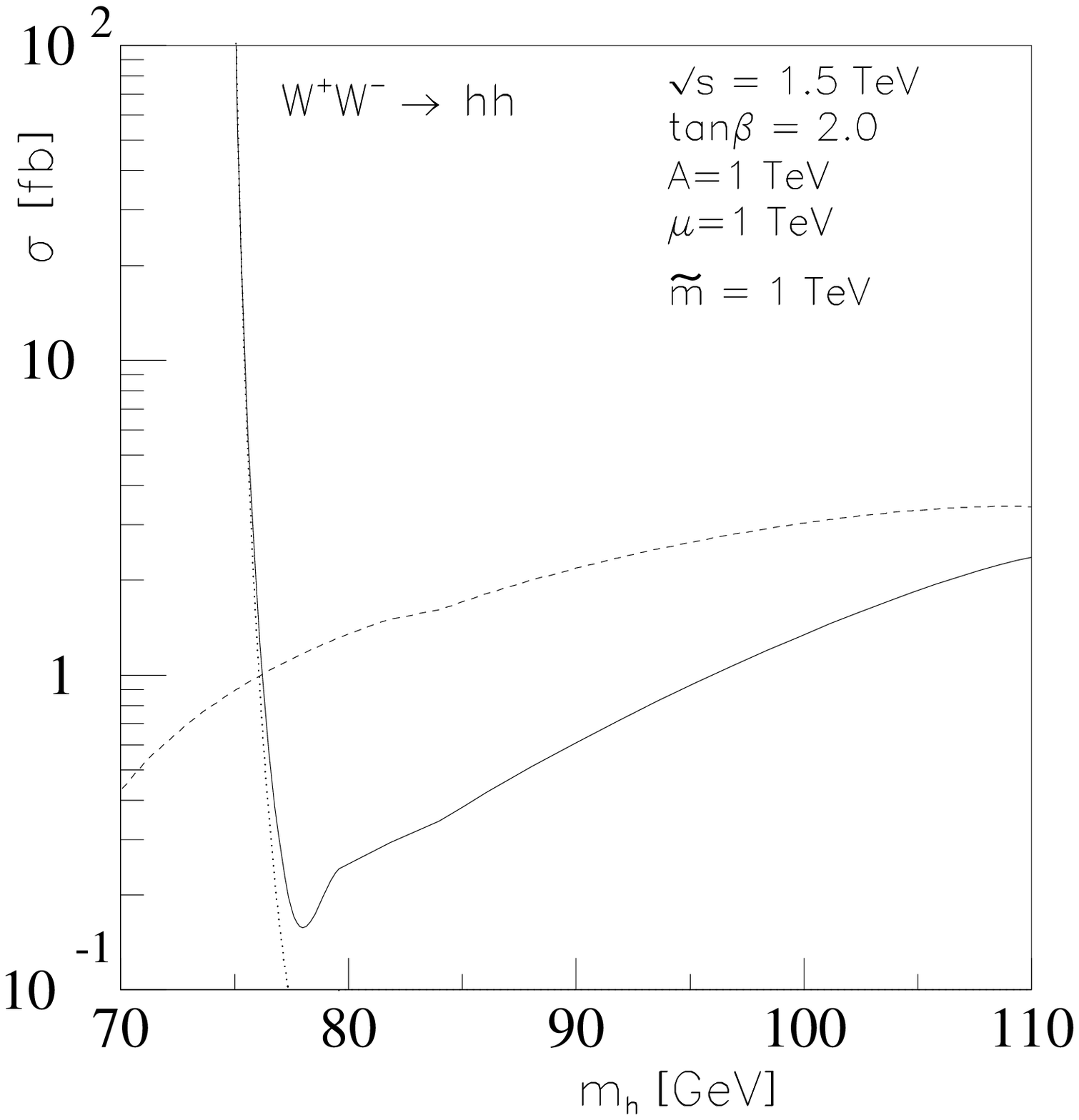}}}
\end{picture}
\vspace*{-12mm}
\caption{Cross section $\sigma(e^+e^-\to \nu_e\bar\nu_e hh)$ 
(via $WW$ fusion) as a function of $m_h$.
The dotted curve is the resonant production,
the dashed curve gives the decoupling limit.
}
\end{center}
\end{figure}

We show in Fig.~\ref{Fig:sig-WW-2} the resulting $WW$ fusion cross section, 
at $\sqrt{s}=1.5~\TeV$, and for $\tilde m = 1~\TeV$.
The structure is reminiscent of Fig.~\ref{Fig:sig-Zll-2},
and the reasons for this are the same. Notice, however, that 
the scale is different. Since this is a fusion cross section,
it grows logarithmically with energy.

For high values of $m_h$ we see that there is a moderate contribution
to the cross section from the non-resonant part.
For a lower squark mass scale $\tilde m$, the situation is 
somewhat different.
In the absence of mixing, the light Higgs particle then tends to be
lighter (for $\tilde m=500~\GeV$, $\tan\beta=2$: $m_h\lsim 90~\GeV$
--- which is mostly ruled out already \cite{ALEPH98}).
With mixing, however, higher Higgs masses can be reached.
\section{Sensitivity to $\lambda_{Hhh}$ and $\lambda_{hhh}$}
We are now ready to combine the results and discuss in which parts
of the $m_A$--$\tan\beta$ plane one might hope to measure the trilinear
couplings $\lambda_{Hhh}$ and $\lambda_{hhh}$.
In Figs.~\ref{Fig:sensi-500-m1000} and \ref{Fig:sensi-500-m500}
we have identified regions according to the following criteria 
\cite{DHZ,OP98}:
\begin{itemize}
\item[(i)]
Regions where $\lambda_{Hhh}$ might become measurable are identified
as those where 
$\sigma(H)\times\mbox{BR}(H\to hh)> 0.1\mbox{ fb}$ (solid),
while simultaneously $0.1 < \mbox{BR}(H\to hh) < 0.9$
[see Figs.~\ref{Fig:BR-H-A}--\ref{Fig:hole-500}].
In view of the recent, more optimistic, view on the
luminosity that might become available, 
we also give the corresponding contours for 0.05~fb (dashed) 
and 0.01~fb (dotted). 
\item[(ii)]
Regions where $\lambda_{hhh}$ might become measurable
are those where the {\it continuum} $WW\to hh$
cross section [Eq.~(\ref{Eq:sigWW-nonres})] is larger than 
0.1~fb (solid).
Also shown are contours at 0.05 (dashed) and 0.01~fb (dotted).
\end{itemize}
We have excluded from the plots the region where $m_h<72.2~\GeV$
\cite{ALEPH98}.
This corresponds to low values of $m_A$ and low $\tan\beta$.

\begin{figure}[htb]
\refstepcounter{figure}
\label{Fig:sensi-500-m1000}
\addtocounter{figure}{-1}
\begin{center}
\setlength{\unitlength}{1cm}
\begin{picture}(12,7.8)
\put(-1.0,1.5)
{\mbox{\epsfysize=7.0cm\epsffile{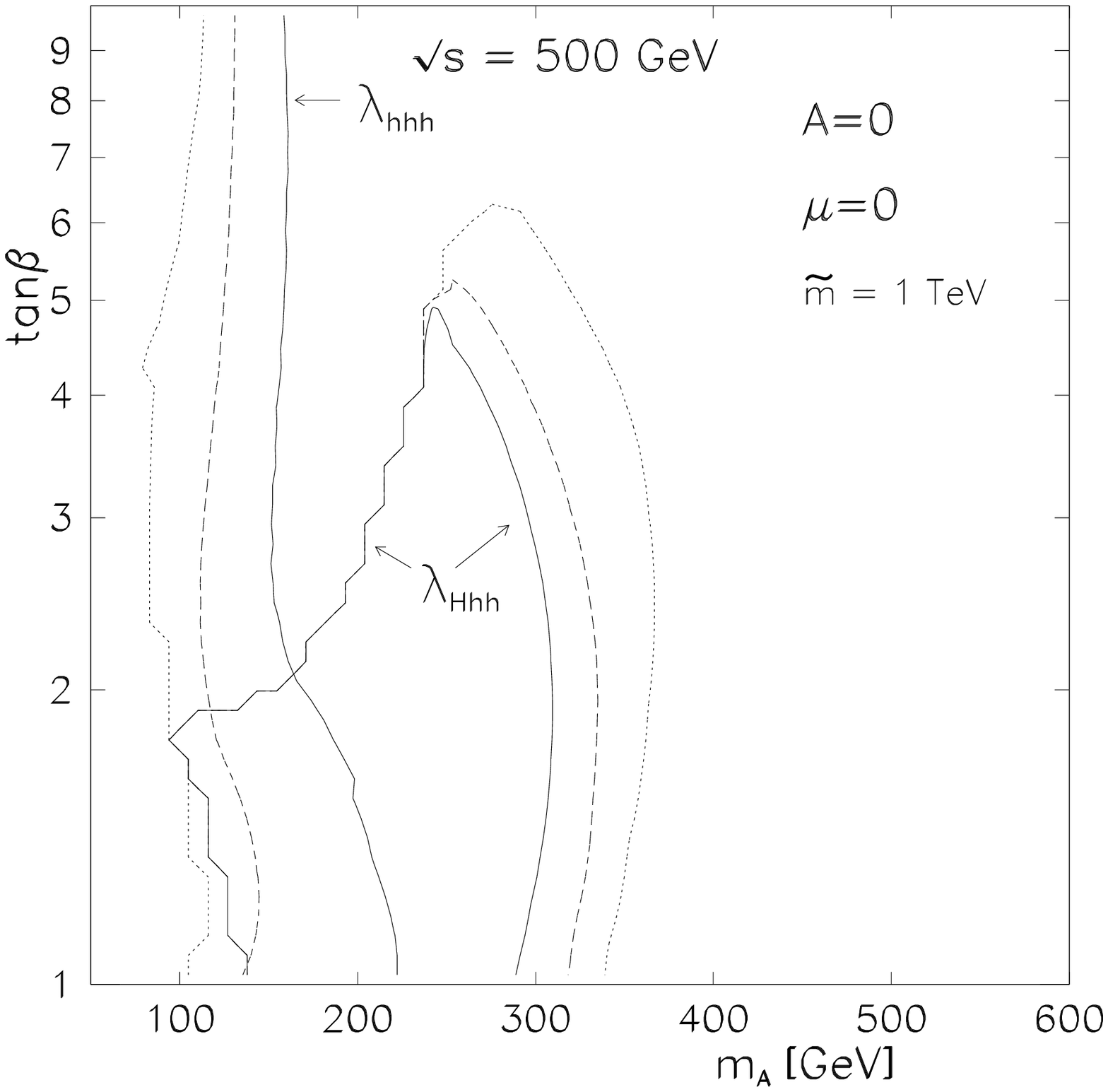}}
 \mbox{\epsfysize=7.0cm\epsffile{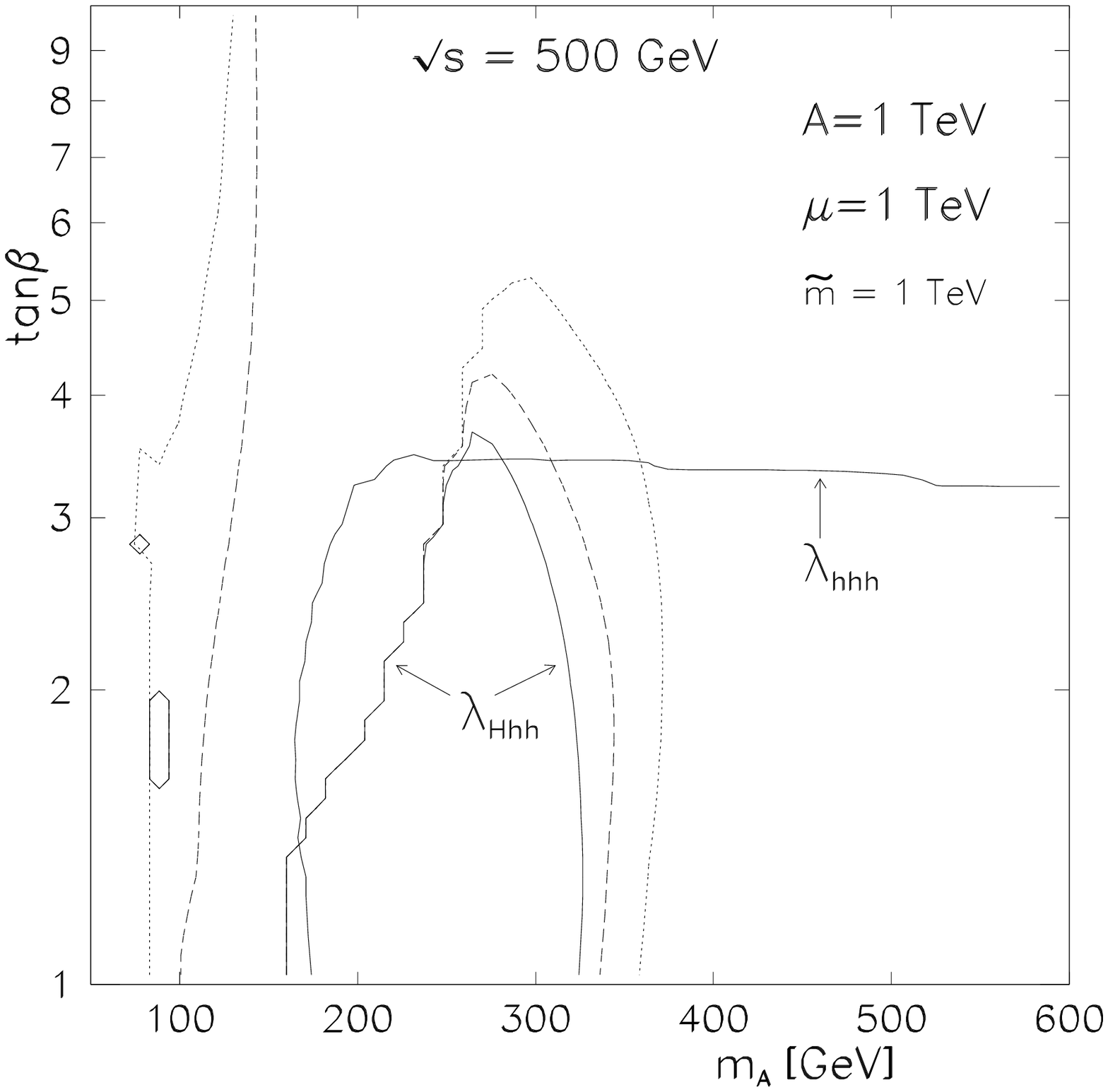}}}
\end{picture}
\vspace*{-12mm}
\caption{Regions where trilinear couplings $\lambda_{Hhh}$ and 
$\lambda_{hhh}$ might be measurable at $\sqrt{s}=500$~GeV.
Inside contours labelled $\lambda_{Hhh}$, 
$\sigma(H)\times\mbox{ BR}(H\to hh) > 0.1~\mbox{fb}$ (solid),
while $0.1<\mbox{BR}(H\to hh)<0.9$.
Inside (to the right or below) contour labelled $\lambda_{hhh}$,
the {\it continuum} $WW\to hh$ cross section exceeds 0.1~fb (solid).
Analogous contours are given for 0.05 (dashed) and 0.01~fb (dotted).
Two cases of squark mixing are considered, as indicated.
}
\end{center}
\end{figure}
\begin{figure}[htb]
\refstepcounter{figure}
\label{Fig:sensi-500-m500}
\addtocounter{figure}{-1}
\begin{center}
\setlength{\unitlength}{1cm}
\begin{picture}(12,7.8)
\put(-1.0,1.5)
{\mbox{\epsfysize=7.0cm\epsffile{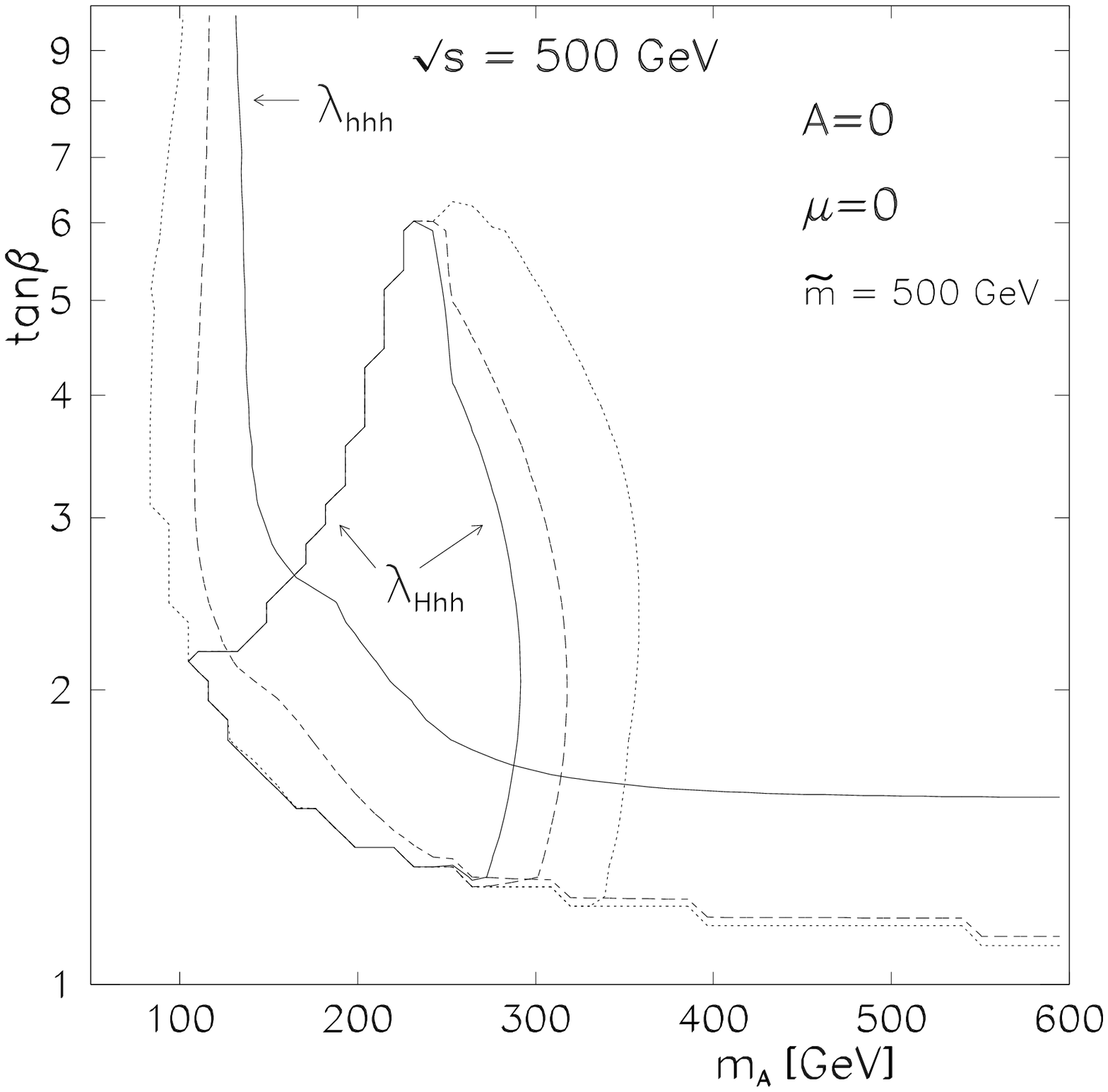}}
 \mbox{\epsfysize=7.0cm\epsffile{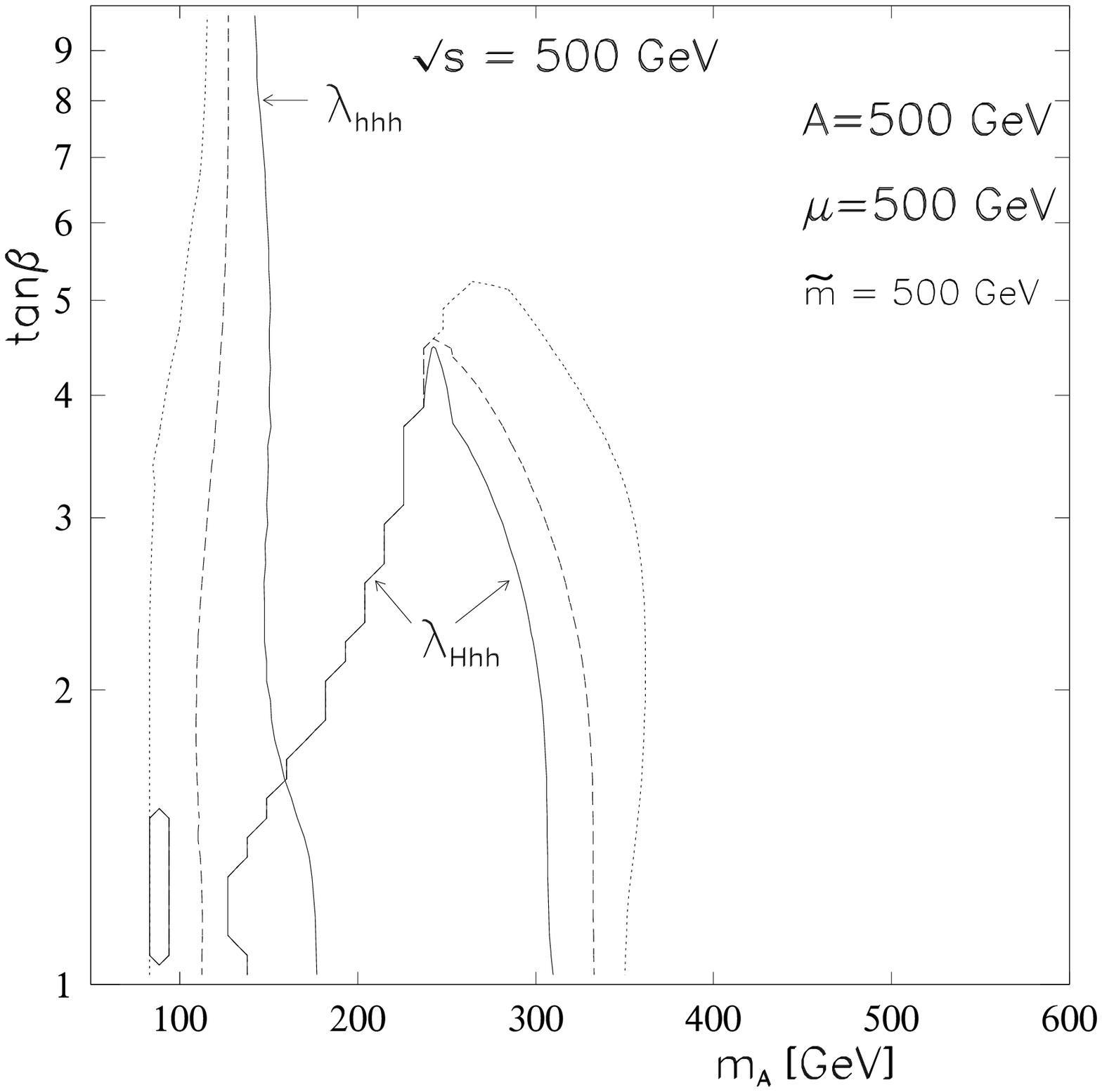}}}
\end{picture}
\vspace*{-12mm}
\caption{Similar to Fig.~\ref{Fig:sensi-500-m1000} for
$\tilde m=500~\GeV$.}
\end{center}
\end{figure}
These cross sections are small, the measurements are not
going to be easy.
With an integrated luminosity of 500~fb$^{-1}$,
the contours at 0.1~fb correspond to 50 events per year.
This will be reduced by efficiencies, but should indicate
the order of magnitude that can be reached.

With increasing luminosity, the region where $\lambda_{Hhh}$ might
be accessible, extends somewhat to higher values of $m_A$.
Note the steep edge around $m_A\simeq200~\GeV$, where increased 
luminosity does not help.
This is determined by the vanishing of $\mbox{BR}(H\to hh)$,
see Figs.~\ref{Fig:hole-1000} and \ref{Fig:hole-500}.

The coupling $\lambda_{hhh}$ is accessible in a much larger part
of this parameter space, but ``large'' values of $\tan\beta$
are accessible only if $A$ is small, or if the luminosity is high.
 
The precise region in the $\tan\beta$--$m_A$ plane, in which these
couplings might be accessible, depends on details of the model.
As a further illustration of this point, 
we show in Fig.~\ref{Fig:sensi-500-m500}
the corresponding plots for a squark mass parameter $\tilde m=500~\GeV$.
In the case of no mixing, there is now a band at small $m_A$ and
small $\tan\beta$ that is excluded by the Higgs mass bound \cite{ALEPH98}.
Furthermore, where the Higgs mass is low, the coupling
$\lambda_{hhh}$ is small [see Fig.~\ref{Fig:lam-mh}], and
a corresponding band is excluded from possible measurements.

\section{$CP$ studies}
In the MSSM, the Higgs sector contains also a particle that is odd
under $CP$. Such particles, 
as well as Higgs-like particles which are not eigenstates of $CP$,
would be expected in more general electroweak theories.
One example would be the two-Higgs-doublet model \cite{TDLee,GHKD}.
Model-independent determinations of the Higgs particle $CP$ are
possible in the Bjorken process as well as in the electron-electron
channel.

There could also be $CP$ violation in the Higgs sector, 
in which case the Higgs bosons would not be $CP$ eigenstates 
\cite{Bernreuther}. 
Such mixing could take place at the tree level
\cite{Deshpande}, or it could be induced by radiative corrections.
It has also been pointed out that such mixing might
take place in the MSSM, and be resonant \cite{Pilaftsis}.
\subsection{The Bjorken process}
Certain distributions for the Bjorken process \cite{Bjorken}
are sensitive to the $CP$ parity.
Suitable observables may also demonstrate presence of $CP$ violation.

Below, we present an effective Lagrangian which contains $CP$ violation 
in the Higgs sector.
$CP$ violation usually appears as a one-loop effect, since
the $CP$-odd coupling introduced below
is a higher-dimensional operator and in renormalizable models these are
induced only at loop level. 
Thus, the effects are expected to be small and the confirmation of 
presence of any $CP$ violation could be rather difficult.

The $ZZh$ coupling is taken to be \cite{NelCha}
\begin{equation}
\label{Eq:coupling}
i2^{5/4}\sqrt{G_F}
\begin{cases}
m_Z^2\,g^{\mu\nu} & \mbox{for }h=H \mbox{ ($CP$ even)}, \\
\eta\, \epsilon^{\mu\nu\rho\sigma} k_{1\rho} k_{2\sigma} &
\mbox{for }h=A \mbox{ ($CP$ odd)},
\end{cases}
\end{equation}
where $k_1$ and $k_2$ are the momenta of the gauge bosons.
The $CP$ odd term originates from the dimension-5 operator 
$\epsilon^{\,\mu \nu \rho \sigma} Z_{\mu \nu} Z_{\rho \sigma} H$.
Simultaneous presence of $CP$-even and $CP$-odd terms leads to $CP$ violation,
whereas presence of only the last term describes a pseudoscalar 
coupling to the vector bosons.

It is well known that the correlation between
the two decay planes spanned by the Dalitz pairs from a $\pi^0$
decay reveal its pseudoscalar nature \cite{YangDal}.
In complete analogy,
the orientation of the decay plane spanned by the momenta
of the fermions from the $Z^0$ which is accompanying the Higgs particle 
in the Bjorken process can be used to determine the $CP$
of the Higgs particle \cite{NelCha,Grzad,SkjOsl95}. 
Other methods have also been suggested \cite{e-p}.
These include studies of correlations among momenta of the initial electron 
and final-state fermions \cite{Arens}.

In fact, a semi-realistic Monte Carlo study shows that (at 300 GeV) 
it should be possible to verify 
the scalar character of the Standard Model Higgs after 
three years of running at a future linear collider \cite{SkjOsl95}.
Also, various ways of searching for $CP$ violation have been suggested
\cite{NelCha,Djouadi,SkjOsl95}. 

\subsection{Electron-electron collisions}
The electron-electron collider mode is interesting since one
may produce states not accessible in the annihilation channel;
also, a large electron polarization will be readily available.
Furthermore, at high energies, the Higgs production at an electron-electron
collider will proceed via gauge boson fusion \cite{Hikasa,Barger}, 
and thus not be suppressed by the $s$-channel annihilation mechanism.
Certain models also predict doubly charged Higgs particles \cite{double-h},
some of which can be produced more readily at an electron-electron collider.

Scalar (``Higgs'') particles, $h$, $h^-$ and $h^{--}$, 
are produced in the $t$-channel via $Z$- or $W$-exchange:
\begin{eqnarray}
\label{eq-e-e-}
e^-(p_1)+e^-(p_2) &\rightarrow& e^-(p'_1)+e^-(p'_2)+h(p_h), \\
e^-(p_1)+e^-(p_2) &\rightarrow& e^-(p'_1)+\nu_e(p'_2)+h^-(p_h), \\
e^-(p_1)+e^-(p_2) &\rightarrow& \nu_e(p'_1)+\nu_e(p'_2)+h^{--}(p_h).
\end{eqnarray}
(In some models, including the left--right symmetric model 
\cite{Mohapatra}, the doubly-charged Higgs boson has practically
no coupling to the ordinary, left-handed $W$ bosons and would not
be produced by this mechanism.)

Several distributions
are quite sensitive to the $CP$ of the Higgs particle.
We see immediately from (\ref{Eq:coupling})
that near the forward direction,
where $\veck_1$ and $\veck_2$ are antiparallel,
the production of a $CP$-odd Higgs boson will be suppressed.

In the $CP$-even case, the Higgs particle tends to be softer, 
and events are more aligned with the beam direction than in 
the $CP$-odd case.
In fact, the Higgs energy distribution may be one of the better
observables for discriminating the two cases, as illustrated 
in Fig.~\ref{Fig:eh}.

\begin{figure}[htb]
\refstepcounter{figure}
\label{Fig:eh}
\addtocounter{figure}{-1}
\phantom{AAA}
\begin{center}
\setlength{\unitlength}{1cm}
\begin{picture}(12,7.2)
\put(-1.0,1.5)
{\mbox{\epsfysize=6.2cm\epsffile{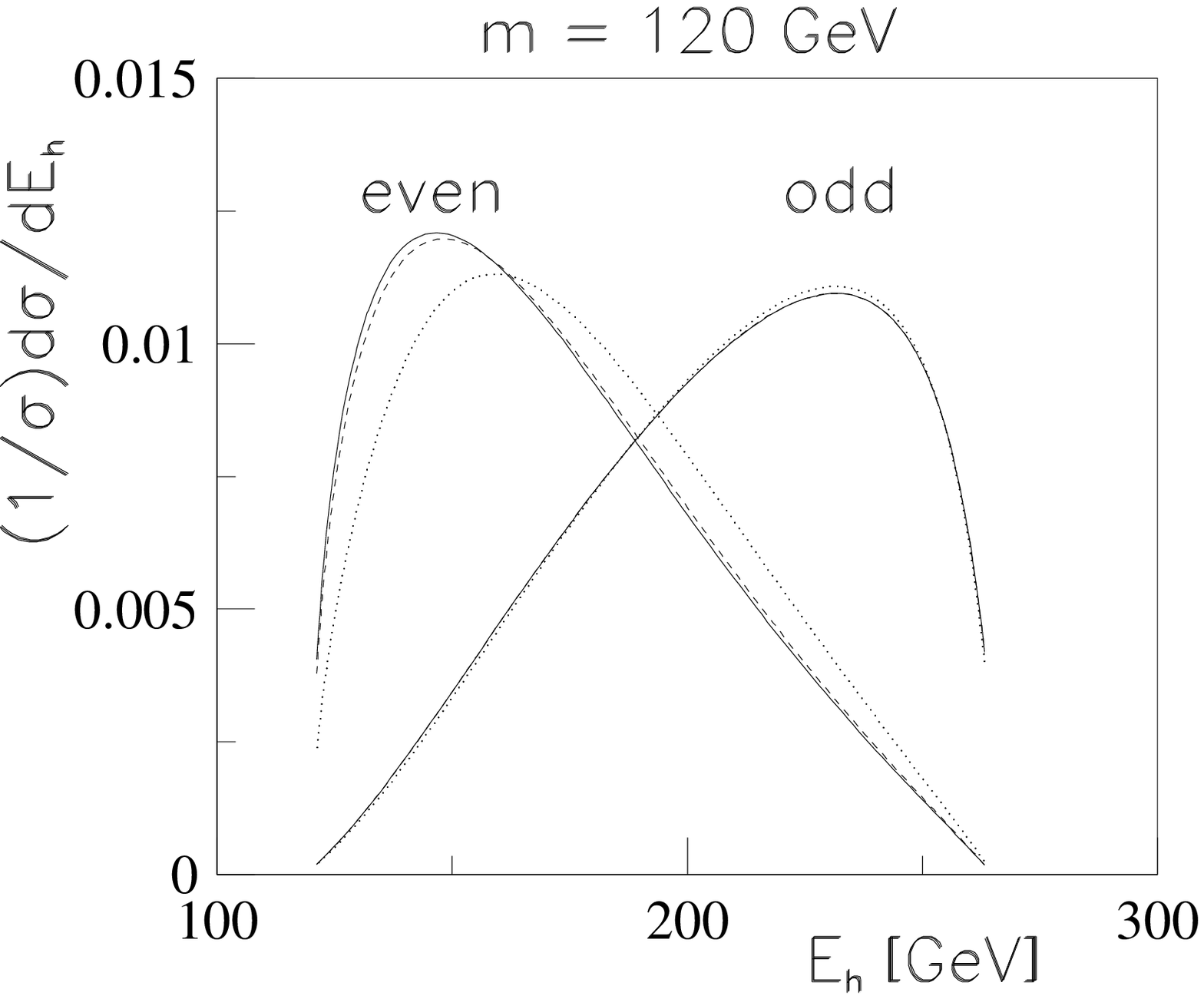}}
 \mbox{\epsfysize=6.2cm\epsffile{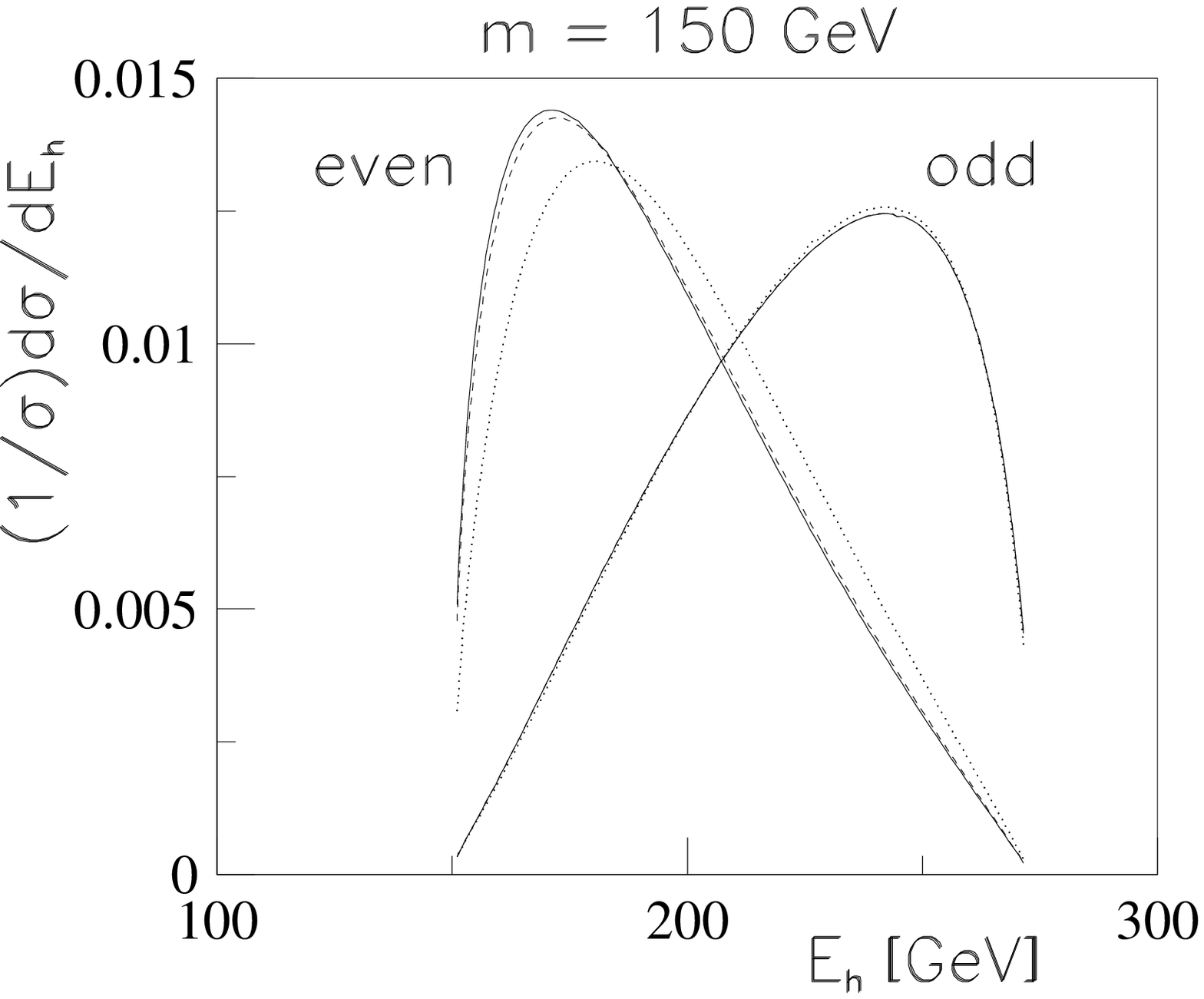}}}
\end{picture}
\end{center}
\vspace*{-2.0cm}
\caption{Higgs energy spectra for the case $E_{\rm c.m.}=500$~GeV,
and for Higgs masses $m_h=120$~GeV and 150~GeV.
The solid curves give the distributions in the absence of any cut.
The dashed and dotted curves show the corresponding distributions when 
cuts at $5\deg$ and $15\deg$ are imposed on the electron momenta.}
\end{figure}

The dependence on longitudinal beam polarization ($P_1$ and $P_2$)
enters in the following way
\begin{equation}
\label{Eq:pol-dep}
\dd^4\sigma^{(h)}
=\dd^4\sigma_0^{(h)}
\left[1 +A_1^{(h)}\, P_1P_2 +A_2^{(h)}(P_1+P_2)\right].
\end{equation}
It turns out that the dependence on the {\it product} of the two
beam polarizations is much larger in the $CP$-odd case.
This dependence, which is represented by the observable 
$A_1$ (see Fig.~\ref{Fig:sigtot-mh-a1}),
becomes a better ``discriminator'' for increasing Higgs masses,
when the Higgs momentum decreases, and other methods therefore tend 
to become less efficient.

\begin{figure}[htb]
\refstepcounter{figure}
\label{Fig:sigtot-mh-a1}
\addtocounter{figure}{-1}
\phantom{AAA}
\begin{center}
\setlength{\unitlength}{1cm}
\begin{picture}(12,8.5)
\put(2.0,2.5)
{\mbox{\epsfysize=6.5cm\epsffile{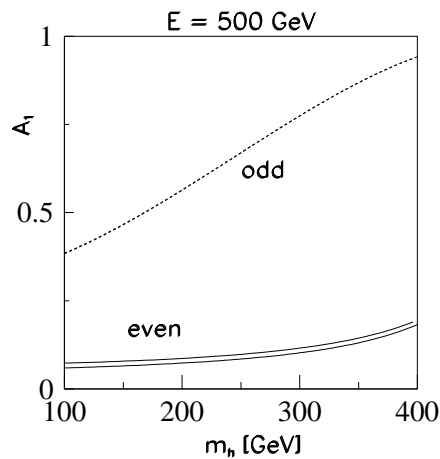}}}
\end{picture}
\end{center}
\vspace*{-3.0cm}
\caption{The bi-polarization-dependence $A_1$ 
[see Eq.~(\ref{Eq:pol-dep})] as obtained from the integrated
cross sections for Higgs production in electron-electron
collisions at $E_{\rm c.m.}=500$~GeV, for a range of Higgs masses.
Standard Model (denoted ``even'') and $CP$-odd results are shown.
For the even case, the lower curve corresponds to no cut,
whereas the upper ones are obtained with an angular cut on the
final-state electron momenta at $10\deg$.
(For the odd case, the two curves are indistinguishable.)}
\end{figure}

If the two final-state electrons are observed, a certain azimuthal
distribution, as well as the electron polar-angle distributions,
will also be useful for discriminating the two cases \cite{Boe}.
There are also ways to search for possible parity-violating effects 
in the $ZZ$-Higgs coupling.

\section{Conclusions}
We have reviewed the results of a detailed investigation \cite{OP98}
of the possibility of measuring
the MSSM trilinear couplings $\lambda_{Hhh}$ and $\lambda_{hhh}$
at an $e^+ e^-$ collider,
focusing on the importance of mixing in the squark sector,
as induced by the trilinear coupling $A$ and the bilinear coupling $\mu$.

At moderate energies ($\sqrt{s}=500~\GeV$) the range in 
the $m_A$--$\tan\beta$ plane that is accessible for studying 
$\lambda_{Hhh}$ changes quantitatively for non-zero values of
the parameters $A$ and $\mu$.
As far as the coupling $\lambda_{hhh}$
is concerned, however, there is a qualitative change from the case of
no mixing in the squark sector.
If $A$ is large, then high luminosity is required, in order to reach 
``high'' values of $\tan\beta$.
At higher energies ($\sqrt{s}=1.5~\TeV$), the mixing parameters
$A$ and $\mu$ change the accessible region of the
parameter space only in a quantitative manner.

We have also given a brief review of some ways to investigate
$CP$ properties of the Higgs particles, in $e^+e^-$ as well
as in $e^-e^-$ collisions.
\medskip

This research was supported by the Research Council of Norway.
It is a pleasure to thank the organizers of the Epiphany conference
for creating a most stimulating meeting.
The topics reported on here have been explored with various 
colleagues whom I would like to thank for stimulating collaborations.
Finally, it is with profound gratitude I acknowledge the inspiration
that has been provided by Bj{\o}rn H.\ Wiik, whose enthusiasm
for basic physics challenges influenced so many of us.


\end{document}